\documentclass[11pt]{scrartcl}
\usepackage[utf8]{inputenc}
\usepackage[T1]{fontenc}
\usepackage{amssymb}
\usepackage{color}
\usepackage{enumerate}

\usepackage{relsize,xspace}
\usepackage[usenames,dvipsnames]{xcolor}
\usepackage{mathtools}

\usepackage{microtype}
\usepackage{amsmath}
\usepackage{amssymb}
\usepackage{amsfonts}
\usepackage{stmaryrd}
\usepackage{tikz}
\usepackage{bm}
\usepackage{fullpage}
\usepackage{todonotes}
\usepackage{mathrsfs}
\usepackage{lmodern}

\usetikzlibrary{calc,through,arrows,arrows.meta,automata,decorations.pathmorphing,decorations.pathreplacing,shapes,backgrounds,fit,positioning,shapes.geometric,patterns}
\usetikzlibrary{decorations.text}
\usetikzlibrary{decorations}
\tikzset{
	position/.style args={#1:#2 from #3}{
		at=($(#3)+(#1:#2)$)
	}
}

\pgfdeclarelayer{bg}    
\pgfsetlayers{bg,main}  

\definecolor{blackblue}{rgb}{0,0.18,0.39}
\usepackage[ocgcolorlinks, allcolors={blackblue}]{hyperref}

\definecolor{magenta}{rgb}{0.79, 0.08, 0.48}
\definecolor{AO}{rgb}{0.0, 0.5, 0.0}
\definecolor{phthaloblue}{rgb}{0.0, 0.06, 0.54}
\definecolor{pistachio}{rgb}{0.58, 0.77, 0.45}
\definecolor{darkgoldenrod}{rgb}{0.72, 0.53, 0.04}

\usepackage[amsmath,thmmarks,hyperref]{ntheorem}
\usepackage{cleveref}

\crefformat{page}{#2page~#1#3}%
\Crefformat{page}{#2Page~#1#3}%
\crefformat{equation}{#2(#1)#3}%
\Crefformat{equation}{#2(#1)#3}%
\crefformat{figure}{#2Figure~#1#3}%
\Crefformat{figure}{#2Figure~#1#3}%
\crefformat{section}{#2Section~#1#3}
\Crefformat{section}{#2Section~#1#3}
\crefformat{chapter}{#2Chapter~#1#3}
\Crefformat{chapter}{#2Chapter~#1#3}
\crefformat{chapter*}{#2Chapter~#1#3}
\Crefformat{chapter*}{#2Chapter~#1#3}
\crefformat{part}{#2Part~#1#3}
\Crefformat{part}{#2Part~#1#3}
\crefformat{enumi}{#2(#1)#3}
\Crefformat{enumi}{#2(#1)#3}

\usepackage{latexsym}

\theoremnumbering{arabic}
\theoremstyle{plain}
\theoremsymbol{}
\theorembodyfont{\itshape}
\theoremheaderfont{\normalfont\bfseries}
\theoremseparator{}

\newtheorem{theorem}{Theorem}[section]
\crefformat{theorem}{#2Theorem~#1#3}
\Crefformat{theorem}{#2Theorem~#1#3}

\newcommand{\newtheoremwithcrefformat}[2]{%
	\newtheorem{#1}[theorem]{#2}%
	\crefformat{#1}{##2\MakeUppercase#1~##1##3}%
	\Crefformat{#1}{##2\MakeUppercase#1~##1##3}%
}

\newtheoremwithcrefformat{proposition}{Proposition}
\newtheoremwithcrefformat{observation}{Observation}
\newtheoremwithcrefformat{lemma}{Lemma}
\newtheoremwithcrefformat{conjecture}{Conjecture}
\newtheoremwithcrefformat{corollary}{Corollary}
\theorembodyfont{\upshape}
\newtheoremwithcrefformat{example}{Example}
\newtheoremwithcrefformat{remark}{Remark}
\newtheoremwithcrefformat{definition}{Definition}

\theoremsymbol{\ensuremath{\square}}

\theoremstyle{nonumberplain}
\theoremheaderfont{\scshape}
\theorembodyfont{\normalfont}
\theoremsymbol{\ensuremath{\square}}
\newtheorem{proof}{Proof.}

\newenvironment{cenv}{\begin{list}{}{%
			\setlength{\labelwidth}{1.5em}%
			\setlength{\leftmargin}{\labelwidth}%
			\addtolength{\leftmargin}{\labelsep}%
			\setlength{\listparindent}{0em}%
			\setlength{\topsep}{10pt}%
			\setlength{\itemsep}{5pt}%
			\setlength{\parsep}{0pt}%
		}
	}{
	\end{list}
}

\newcounter{claimcounter}
\newenvironment{Claim}{
	
	\refstepcounter{claimcounter}
	\begin{cenv}
		\item[{Claim \arabic{claimcounter}.}]
	}{
	\end{cenv}
}

\newenvironment{ClaimProof}[1][]{\noindent{%
		\ifthenelse{\equal{#1}{}}{\textsl{Proof.\ }}{\textsl{#1.\ }}%
}}{\hspace*{1em}\nobreak\hfill$\dashv$\endtrivlist\addvspace{2ex plus
		0.5ex minus0.1ex}}

\newcounter{algorithmcounter}
\newenvironment{algorithm}{
	
	~\refstepcounter{algorithmcounter}
	\begin{cenv}
		\item[{\textbf{Algorithm \arabic{algorithmcounter}.}}]
	}{
	\end{cenv}
}

\newcommand{\N}{\mathbb{N}}

\newcommand{\Start}[1]{\mathscr{S}(#1)}
\newcommand{\End}[1]{\mathscr{E}(#1)}
\newcommand{\sizeof}[1]{\left|#1\right|}

\newcommand{\updatesequence}{\sigma}

\newcommand{\touches}{\succ}

\newcommand{\BTSsets}{\operatorname{PermutList}}
\newcommand{\BTSset}[1]{\operatorname{\BTSsets}(#1)}
\newcommand{\BTS}[1]{\operatorname{TouchSeq}(#1)}
\newcommand{\TBs}[1]{\operatorname{TouchingBlocks}(#1)}
\newcommand{\consistent}{\approx}

\newcommand{\Label}{\operatorname{Label}}
\newcommand{\Labels}[1]{\operatorname{Labels}(#1)}
\newcommand{\blockset}{\mathcal{B}}

\title{Congestion-Free Rerouting of Flows on DAGs\footnote{The
		research of Saeed Akhoondian Amiri and Sebastian Wiederrecht has been supported by 
		the European Research Council (ERC) under the European Union's Horizon
		2020 research and innovation programme (ERC consolidator grant DISTRUCT,
		agreement No 648527). Stefan Schmid is supported by the Danish VILLUM foundation project 
		\emph{ReNet}.}}

\author{
	Saeed Akhoondian Amiri\thanks{TU Berlin, DE \texttt{saeed.amiri@tu-berlin.de}}
	\quad Szymon Dudycz\thanks{University of Wroclaw, PL \texttt{szymon.dudycz@gmail.com}} \\ Stefan
	Schmid\thanks{Aalborg University, DK \texttt{schmiste@cs.aau.dk}}
	\quad Sebastian Wiederrecht\thanks{TU Berlin, DE \texttt{sebastian.wiederrecht@tu-berlin.de}}}

\begin{document}
	
	\date{}
	
	\maketitle
	
	\sloppy
	
	\begin{abstract}
		Changing a given configuration in a graph into 
		another one is known as a reconfiguration problem. Such 
		problems have recently received much interest in the context of algorithmic graph theory.
		We initiate the theoretical study of the following reconfiguration problem: 
		How to reroute $k$ unsplittable
		flows of a certain demand in a capacitated network
		from their
		current paths to their respective new paths,
		in a congestion-free manner?
		This problem finds immediate 
		applications, e.g., in traffic engineering in computer networks.
		We show that the problem is generally NP-hard already for
		$k=2$ flows, which motivates us to study rerouting on a 
		most basic class of 
		flow graphs, 
		namely DAGs. Interestingly, we find that for general $k$, deciding
		whether an unsplittable multi-commodity flow rerouting schedule
		exists, is NP-hard even on DAGs. Both NP-hardness proofs are non-trivial.
		Our main contribution is a polynomial-time (fixed parameter tractable)
		algorithm
		to solve the route update problem for a 
		bounded number of flows on DAGs. At the heart of our algorithm lies a novel decomposition 
		of the flow network that allows us to express and resolve reconfiguration
		dependencies among flows.	
	\end{abstract}
	
	
	\section{Introduction}\label{sec:intro}
	
	Reconfiguration problems are combinatorial problems 
	which ask for a transformation of one configuration into
	another one, subject to some (reconfiguration) rules.
	Reconfiguration problems are fundamental and have been
	studied in many contexts, 
	including puzzles and games (such as
	Rubik's cube)~\cite{van2013complexity}, satisfiability~\cite{gopalan2009connectivity}, 
	independent sets~\cite{hearn2005pspace}, 
	vertex coloring~\cite{cereceda2011finding}, or
	matroid bases~\cite{ito2008complexity}, to just name a few. 
	
	Reconfiguration problems also naturally arise in 
	the context of networking applications and routing.
	For example, a fundamental problem in computer
	networking regards the question of how to 
	reroute traffic from the current path $p_1$ to a given new 
	path $p_2$, by changing the forwarding rules at routers
	(the \emph{vertices}) one-by-one, while maintaining
	certain properties  \emph{during} the reconfiguration 
	(e.g., short path lengths~\cite{bonsma2013complexity}).  
	Route reconfigurations (or \emph{updates}) are frequent 
	in computer networks: paths are changed, e.g., to account for changes
	in the security policies,
	in response to new route advertisements, 
	during maintenance (e.g., replacing a router),
	to support the migration of virtual machines, etc.~\cite{update-survey}.
	
	This paper initiates the study of a basic 
	\emph{multi-commodity flow rerouting problem}:
	how to reroute a set of \emph{unsplittable flows} 
	(with certain bandwidth demands) in a capacitated
	network, from
	their current paths to their respective new paths
	\emph{in a congestion-free manner}. The problem finds
	immediate applications in traffic engineering~\cite{rsvp}, whose 
	main objective is to avoid network congestion. 
	Interestingly, while congestion-aware routing  
	and traffic engineering problems have
	been studied
	intensively in the 
	past~\cite{AmiriKMR16,ChekuriEP16,chekuri2015multicommodity,shimon-flows,KawarabayashiKK14,ufkleinberg,leighton1999multicommodity,ufskutella},
	surprisingly little is known today
	about the problem of how to reconfigure resp.~\emph{update} the routes of flows. 
	Only recently, due to the advent of
	Software-Defined Networks (SDNs), the problem has received
	much attention in the networking 
	community~\cite{sirocco16update,roger-infocom,Forster2016Consistent,ludwig2015scheduling}.
	
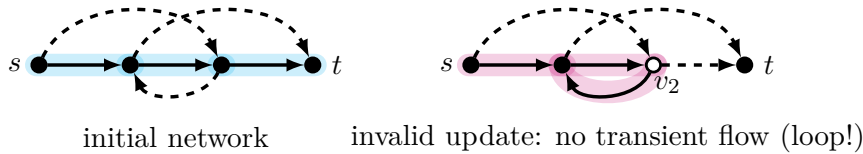
\begin{figure}[h!]
		\begin{center}
			\begin{tikzpicture}[scale=0.8]
			\tikzset{>=latex} 
			\tikzstyle{main} = [inner sep=0pt,minimum size=7pt,draw,circle,fill,thick,scale=0.8]
			\tikzstyle{ghost} = [inner sep=0pt,scale=0.85]
			\tikzstyle{flow} = [line width=8.5pt,line cap=round,opacity=0.2]
			\tikzstyle{active} = [line width=1.3pt,->]
			\tikzstyle{inactive} = [line width=1.3pt,->,dashed]
			
			\node (o) [] {};	
			
			\node (c-1) [position=180:13mm from o] {};
			\node (c-2) [position=0:13mm from o] {};
			
			\node (1-4) [main,position=0:0mm from c-1] {};
			\node (1-3) [main,position=180:15mm from 1-4] {};
			\node (1-2) [main,position=180:15mm from 1-3] {};
			\node (1-1) [main,position=180:15mm from 1-2] {};
			
			\node (s1) [position=180:4mm from 1-1] {$s$};
			\node (t1) [position=0:4mm from 1-4] {$t$};
			
			\node (lg1) [ghost,position=180:7.5mm from 1-3] {};
			\node (l1) [position=270:12mm from lg1] {initial network};
			
			\node(g1-1) [ghost,position=180:0.7mm from 1-1] {};
			\node(g1-4) [ghost,position=0:0.7mm from 1-4] {};
			\node(g1-2) [ghost,position=0:0.7mm from 1-2] {};
			\node(g1-3) [ghost,position=0:0.7mm from 1-3] {};
			\node(g1-2l) [ghost,position=180:0.7mm from 1-2] {};
			\node(g1-3l) [ghost,position=180:0.7mm from 1-3] {};

			\draw (1-1) edge[active] (1-2);
			\draw (1-2) edge[active] (1-3);
			\draw (1-3) edge[active] (1-4);
			
			\draw (1-1) edge[inactive,bend left=65] (1-3);
			\draw (1-3) edge[inactive,bend left=65] (1-2);
			\draw (1-2) edge[inactive,bend left=65] (1-4);

			\node (2-1) [main,position=0:0mm from c-2] {};
			\node (2-2) [main,position=0:15mm from 2-1] {};
			\node (2-3) [main,position=0:15mm from 2-2] {};
			\node (inner2-3) [inner sep=0pt,minimum 
			size=7pt,draw,circle,fill,color=white,scale=0.55,position=0:0mm from 2-3] {};
			\node (2-4) [main,position=0:15mm from 2-3] {};
			
			\node (ul2) [position=305:4mm from 2-3] {$v_2$};
			\node (s2) [position=180:4mm from 2-1] {$s$};
			\node (t2) [position=0:4mm from 2-4] {$t$};
			
			\node (lg2) [ghost,position=180:7.5mm from 2-3] {};
			\node (l2) [position=270:12mm from lg2] {invalid update: no transient flow (loop!)};
			
			\node(g2-1) [ghost,position=180:0.7mm from 2-1] {};
			\node(g2-2u) [ghost,position=135:0.5mm from 2-2] {};
			\node(g2-2) [ghost,position=180:0.7mm from 2-2] {};
			\node(g2-3) [ghost,position=180:0.7mm from 2-3] {};
			\node(g2-3u) [ghost,position=45:0.5mm from 2-3] {};
			\node(g2-2l) [ghost,position=0:0.7mm from 2-2] {};
			\node(g2-3l) [ghost,position=0:0.7mm from 2-3] {};

			\draw (2-1) edge[active] (2-2);
			\draw (2-2) edge[active] (2-3);
			\draw (2-3) edge[inactive] (2-4);
			
			\draw (2-1) edge[inactive,bend left=65] (2-3);
			\draw (2-3) edge[active,bend left=65] (2-2);
			\draw (2-2) edge[inactive,bend left=65] (2-4);

			\begin{pgfonlayer}{bg}    
			
			\draw (g1-1) edge[flow,cyan] (g1-2);
			\draw (g1-2l) edge[flow,cyan] (g1-3);
			\draw (g1-3l) edge[flow,cyan] (g1-4);
			
			\draw (g2-1) edge[flow,magenta] (g2-2l);
			\draw (g2-2) edge[flow,magenta] (g2-3l);
			\draw (g2-3u) edge[flow,magenta,bend left=85] (g2-2u);
			
			\end{pgfonlayer}

			\end{tikzpicture}	
		\end{center}
		\vspace{-1em}
		\caption{\emph{Example:} We are given an initial network consisting of exactly one 
			active flow $F^o$ (solid edges) and the inactive edges (i.e., inactive forwarding rules)
			of the new flow $F^u$ to which we want to 
			reroute (dashed edges). Together we call the two flows an (update) pair $P=(F^o,F^u)$. 
			Updating the outgoing edges of a vertex 
			means activating 
			all previously inactive outgoing edges of $F^u$, and deactivating all other edges of the old 
			flow $F^o$. 
			Initially, the blue flow is a valid (transient) $(s,t)$-flow.
			If the update of vertex $v_2$ takes 
			effect first, an invalid (not transient) flow is introduced (in pink): traffic is forwarded in a 
			loop, 
			hence (temporarily) invalidating
			the path from $s$ to $t$. 
		} 
		\label{fig:ex1}
	\end{figure}

\begin{figure}[h!]
		\begin{center}
			\begin{tikzpicture}[scale=0.8]
			\tikzset{>=latex} 
			\tikzstyle{main} = [inner sep=0pt,minimum size=7pt,draw,circle,fill,thick,scale=0.8]
			\tikzstyle{ghost} = [inner sep=0pt,scale=0.85]
			\tikzstyle{flow} = [line width=8.5pt,line cap=round,opacity=0.2]
			\tikzstyle{active} = [line width=1.3pt,->]
			\tikzstyle{inactive} = [line width=1.3pt,->,dashed]
			
			\node (o) [] {};
			\node (u) [position=270:30mm from o] {};

			\node (c-1) [position=180:13mm from o] {};
			\node (c-2) [position=0:13mm from o] {};
			\node (c-3) [position=180:13mm from u] {};
			\node (c-4) [position=0:13mm from u] {};
			
			\node (1-4) [main,position=0:0mm from c-1] {};
			\node (1-3) [main,position=180:15mm from 1-4] {};
			\node (1-2) [main,position=180:15mm from 1-3] {};
			\node (1-1) [main,position=180:15mm from 1-2] {};
			
			\node (s1) [position=180:4mm from 1-1] {$s$};
			\node (t1) [position=0:4mm from 1-4] {$t$};
			
			\node (lg1) [ghost,position=180:7.5mm from 1-3] {};
			\node (l1) [position=270:12mm from lg1] {initial network};
			
			\node(g1-1) [ghost,position=180:0.7mm from 1-1] {};
			\node(g1-4) [ghost,position=0:0.7mm from 1-4] {};
			\node(g1-2) [ghost,position=0:0.7mm from 1-2] {};
			\node(g1-3) [ghost,position=0:0.7mm from 1-3] {};
			\node(g1-2l) [ghost,position=180:0.7mm from 1-2] {};
			\node(g1-3l) [ghost,position=180:0.7mm from 1-3] {};

			\draw (1-1) edge[active] (1-2);
			\draw (1-2) edge[active] (1-3);
			\draw (1-3) edge[active] (1-4);
			
			\draw (1-1) edge[inactive,bend left=65] (1-3);
			\draw (1-3) edge[inactive,bend left=65] (1-2);
			\draw (1-2) edge[inactive,bend left=65] (1-4);

			\node (2-1) [main,position=0:0mm from c-2] {};
			\node (2-2) [main,position=0:15mm from 2-1] {};
			\node (2-3) [main,position=0:15mm from 2-2] {};
			\node (inner2-2) [inner sep=0pt,minimum 
			size=7pt,draw,circle,fill,color=white,scale=0.55,position=0:0mm from 2-2] {};
			\node (2-4) [main,position=0:15mm from 2-3] {};
			
			\node (s2) [position=180:4mm from 2-1] {$s$};
			\node (t2) [position=0:4mm from 2-4] {$t$};
			
			\node (lg2) [ghost,position=180:7.5mm from 2-3] {};
			\node (l2) [position=270:12mm from lg2] {first update};
			
			\node(g2-1) [ghost,position=180:0.7mm from 2-1] {};
			\node(g2-2u) [ghost,position=135:0.5mm from 2-2] {};
			\node(g2-2a) [ghost,position=225:0.5mm from 2-2] {};
			\node(g2-2) [ghost,position=180:0.7mm from 2-2] {};
			\node(g2-3) [ghost,position=180:0.7mm from 2-3] {};
			\node(g2-3u) [ghost,position=45:0.5mm from 2-3] {};
			\node(g2-2l) [ghost,position=0:0.7mm from 2-2] {};
			\node(g2-3l) [ghost,position=0:0.7mm from 2-3] {};
			\node(g2-4a) [ghost,position=315:0.5mm from 2-4] {};

			\draw (2-1) edge[active] (2-2);
			\draw (2-2) edge[inactive] (2-3);
			\draw (2-3) edge[active] (2-4);
			
			\draw (2-1) edge[inactive,bend left=65] (2-3);
			\draw (2-3) edge[inactive,bend left=65] (2-2);
			\draw (2-2) edge[active,bend left=65] (2-4);

			\node (3-4) [main,position=0:0mm from c-3] {};
			\node (3-3) [main,position=180:15mm from 3-4] {};
			\node (3-2) [main,position=180:15mm from 3-3] {};
			\node (inner3-3) [inner sep=0pt,minimum 
			size=7pt,draw,circle,fill,color=white,scale=0.55,position=0:0mm from 3-3] {};
			\node (3-1) [main,position=180:15mm from 3-2] {};
			
			\node (s3) [position=180:4mm from 3-1] {$s$};
			\node (t3) [position=0:4mm from 3-4] {$t$};
			
			\node (lg3) [ghost,position=180:7.5mm from 3-3] {};
			\node (l3) [position=270:12mm from lg3] {second update};
			
			\node(g3-1) [ghost,position=180:0.7mm from 3-1] {};
			\node(g3-2u) [ghost,position=135:0.5mm from 3-2] {};
			\node(g3-2a) [ghost,position=225:0.5mm from 3-2] {};
			\node(g3-2) [ghost,position=180:0.7mm from 3-2] {};
			\node(g3-3) [ghost,position=180:0.7mm from 3-3] {};
			\node(g3-3u) [ghost,position=45:0.5mm from 3-3] {};
			\node(g3-2l) [ghost,position=0:0.7mm from 3-2] {};
			\node(g3-3l) [ghost,position=0:0.7mm from 3-3] {};
			\node(g3-4a) [ghost,position=315:0.5mm from 3-4] {};

			\draw (3-1) edge[active] (3-2);
			\draw (3-2) edge[inactive] (3-3);
			\draw (3-3) edge[inactive] (3-4);
			
			\draw (3-1) edge[inactive,bend left=65] (3-3);
			\draw (3-3) edge[active,bend left=65] (3-2);
			\draw (3-2) edge[active,bend left=65] (3-4);

			\node (4-1) [main,position=0:0mm from c-4] {};
			\node (4-2) [main,position=0:15mm from 4-1] {};
			\node (4-3) [main,position=0:15mm from 4-2] {};
			\node (4-4) [main,position=0:15mm from 4-3] {};
			\node (inner4-1) [inner sep=0pt,minimum 
			size=7pt,draw,circle,fill,color=white,scale=0.55,position=0:0mm from 4-1] {};
			
			\node (s4) [position=180:4mm from 4-1] {$s$};
			\node (t4) [position=0:4mm from 4-4] {$t$};
			
			\node (lg4) [ghost,position=180:7.5mm from 4-3] {};
			\node (l4) [position=270:12mm from lg4] {final update};
			
			\node(g4-1) [ghost,position=180:0.7mm from 4-1] {};
			\node(g4-2u) [ghost,position=135:0.5mm from 4-2] {};
			\node(g4-2a) [ghost,position=225:0.5mm from 4-2] {};
			\node(g4-2) [ghost,position=180:0.7mm from 4-2] {};
			\node(g4-3) [ghost,position=180:0.7mm from 4-3] {};
			\node(g4-3u) [ghost,position=45:0.5mm from 4-3] {};
			\node(g4-2l) [ghost,position=0:0.7mm from 4-2] {};
			\node(g4-3l) [ghost,position=0:0.7mm from 4-3] {};
			\node(g4-4a) [ghost,position=315:0.5mm from 4-4] {};
			\node(g4-1a) [ghost,position=225:0.5mm from 4-1] {};
			\node(g4-3a) [ghost,position=315:0.5mm from 4-3] {};

			\draw (4-1) edge[inactive] (4-2);
			\draw (4-2) edge[inactive] (4-3);
			\draw (4-3) edge[inactive] (4-4);
			
			\draw (4-1) edge[active,bend left=65] (4-3);
			\draw (4-3) edge[active,bend left=65] (4-2);
			\draw (4-2) edge[active,bend left=65] (4-4);

			\node (ul2) [position=235:4mm from 2-2] {$v_1$};
			\node (ul3) [position=305:4mm from 3-3] {$v_2$};

			\begin{pgfonlayer}{bg}    
			
			\draw (g1-1) edge[flow,cyan] (g1-2);
			\draw (g1-2l) edge[flow,cyan] (g1-3);
			\draw (g1-3l) edge[flow,cyan] (g1-4);
			
			\draw (g2-1) edge[flow,cyan] (g2-2l);
			\draw (g2-2a) edge[flow,cyan,bend left=85] (g2-4a);
			
			\draw (g3-1) edge[flow,cyan] (g3-2l);
			\draw (g3-2a) edge[flow,cyan,bend left=85] (g3-4a);
			
			\draw (g4-1a) edge[flow,cyan,bend left=85] (g4-3a);
			\draw (g4-2a) edge[flow,cyan,bend left=85] (g4-4a);
			\draw (g4-3u) edge[flow,cyan,bend left=85] (g4-2u);
			
			\end{pgfonlayer}
			\end{tikzpicture}	
		\end{center}
		\vspace{-1em}
		\caption{\emph{Example:} We revisit the network of Figure~\ref{fig:ex1} 
			and reroute from $F^o$ to $F^u$ without interrupting the connection 
			between $s$ and $t$ 
			along a unique (transient) path (in blue). To avoid the problem seen in Figure~\ref{fig:ex1}, 
			we first update the vertex $v_2$ in order to establish a shorter connection from $s$ to $t$.
			Once this update has been performed, the update of $v_2$ can be performed without 
			creating a loop. Finally, 	
			by updating $s$, 
			we complete the rerouting.
		} 
		\label{fig:ex2}
	\end{figure}
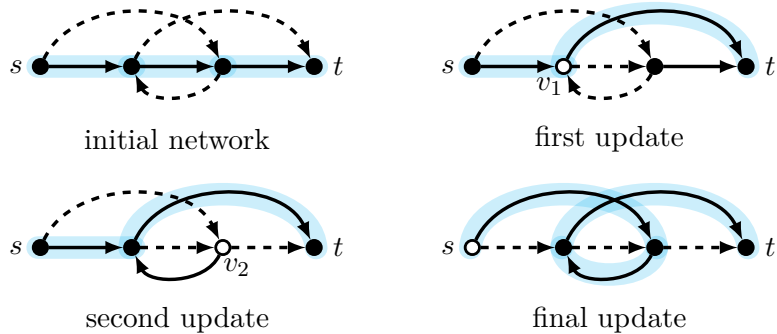

	Figure~\ref{fig:ex1} presents a simple example of the 
	consistent rerouting problem considered in this paper,
	for just a \emph{single} flow: the flow needs to be rerouted
	from the solid path to the dashed path, by changing the forwarding
	links at routers one-by-one.
	The example illustrates a problem that might arise from 
	updating the vertices in an invalid order:  if vertex $v_2$ is updated
	first, a forwarding loop is introduced:
	the transient flow from $s$ to $t$ becomes invalid. 
	Thus, router updates need to be scheduled intelligently over time: 
	A feasible sequence of updates for this example is given in 
	Figure~\ref{fig:ex2}. 
	Note that the example is kept simple intentionally: 
	when moving from a single flow to multiple flows, 
	additional challenges are introduced, as the flows may 
	compete for bandwidth
	and hence interfere.
	We will later discuss a more detailed example, demonstrating a congestion-free
	update schedule for multiple flows. 
	
	\noindent\textbf{Contributions.}
	This paper initiates the algorithmic study of a fundamental 
	unsplittable multicommodity flow rerouting problem. We present a rigorous formal model
	and show that the problem of rerouting
	flows in a congestion-free manner is NP-hard already for two flows
	on general graphs. This motivates us to focus on a 
	most fundamental type of flow graphs, namely the DAG. 
	The main results presented in this paper are the following:
	\begin{enumerate}
		\item Deciding whether a consistent
		network update schedule exists in general graphs is NP-hard, 
		already for 2 flows. 
		\item For general $k$, deciding whether a feasible schedule exists is 
		NP-hard even on loop-free networks (i.e., DAGs). 
		\item For constant $k$, we present an elegant linear-time (fixed parameter tractable)
		algorithm which (deterministically) finds a feasible update schedule on DAGs in time and 
		space $2^{O(k\log k)}O(\sizeof{G})$,
		whenever such a consistent update schedule exists.
	\end{enumerate}
	
	Against the backdrop that the problem of \emph{routing} disjoint paths on DAGs is known to be 
	$W[1]$-hard~\cite{slivkins2010parameterized}
	and finding routes \emph{subject to congestion} even
        harder~\cite{AmiriKMR16}, the finding 
	that 
	the multicommodity flow \emph{rerouting} problem is fixed parameter tractable on DAGs is 
	intriguing.

	\noindent\textbf{Technical Novelty.}
	Our algorithm is based on a novel decomposition of the flow graph into
	so-called \emph{blocks}. This block decomposition allows us to express
	dependencies between flows. 
	In principle, up to $k$ flows (of unit capacity) can share a physical link of 
	capacity $k$, and hence, dependencies arise not between pairs
	but between entire \emph{subsets} of flows along the paths, potentially rendering
	the problem combinatorial: For every given node, there are up to $k!$ possible 
	flow update orders, leading to a brute
	force complexity of $O(k!^{\sizeof{G}})$. However, using a 
	sequence of lemmas,
	we (1) leverage our block decomposition approach, (2) observe
	that many of dependencies are redundant,
	and (3) linearize dependencies, to eventually
	construct a polynomial-sized graph: this graph has the property that
	its independent sets characterize dependencies of 
	the block decomposition. 
	We show that this graph is of \emph{bounded path-width}, allowing us to 
	efficiently compute independent sets (using
	standard dynamic programming), and eventually, construct 
	a feasible update schedule. 
	Overall, this results in an algorithm with linear time complexity in
	the graph size $\sizeof{G}=\sizeof{V(G)}+\sizeof{E(G)}$. 
	
	In addition to our algorithmic contributions, we present rigorous NP-hardness 
	proofs which are based on non-trivial insights into the flow rerouting problem.

	\section{Model and Definitions}
	
	Our problem can be described in terms of edge capacitated directed graphs. 
	In what follows, we will assume basic familiarity with directed graphs and we refer the
	reader to~\cite{digraphs} for more background. We denote a directed
	edge $e$ with head $v$ and tail $u$ by $e=(u,v)$. For an undirected
	edge $e$ between vertices $u,v$, we write $e=\{u,v\}$; 
	$u,v$ are called
	endpoints of $e$.
	
	A \textbf{flow network} is a directed uncapacitated graph $G=(V,E,s,t,c)$, 
	where $s$ is the \emph{source}, $t$ the \emph{terminal}, $V$ is the set 
	of vertices with $s,t\in V$, $E\subseteq V\times V$ is a set of ordered 
	pairs known as edges, and $c\colon E\rightarrow\N$ a 
	capacity function assigning a capacity $c(e)$ to every edge $e\in E$.
	
	Our problem, as described above is a multi-commodity flow problem and thus 
	may have \emph{multiple} 
	source-terminal pairs.
	To simplify the notation but without loss of generality,
	in what follows, we define flow networks to have exactly one source and one terminal.
	In fact, we can model any number of different sources and terminals by adding one super source 
	with edges of unlimited capacity to all original sources, and one super terminal with edges of 
	unlimited 
	capacity leading there from all original terminals.
	
	An \emph{$(s,t)$-flow} $F$ of capacity $d\in\N$ is a 
	\emph{directed path} from $s$ to $t$ in a flow network such that 
	$d\leq c(e)$ for all $e\in E(F)$. Given a family 
	$\mathcal{F}$ of $(s,t)$-flows $F_1,\dots,F_k$ with demands 
	$d_1,\dots,d_k$ respectively, we call $\mathcal{F}$ a \textbf{valid flow
		set}, or simply \textbf{valid}, if $c(e)\geq\sum_{i\colon e\in E(F_i)}d_i$.

	Recall that we consider the problem of how to reroute a current (old) flow to a 
	new (update) flow, and hence we will consider such flows in ``update pairs'': 
	
	An \textbf{update flow pair} $P=(F^o,F^u)$ consists 
	of two $(s,t)$-flows $F^o$, the \emph{old flow}, and $F^u$, 
	the \emph{update flow}, each of demand $d$.
	
	A graph $G=(V,E,\mathcal{P},s,t,c)$, where $(V,E,s,t,c)$ is a flow network,
	and $\mathcal{P}=\left\{ P_1,\dots,P_k\right\}$ with
	$P_i=(F^o_i,F^u_i)$, a family of update flow pairs of demand $d_i$, 
	$V=\bigcup_{i\in[k]}V(F^o_i\cup F^u_i)$ and
	$E=\bigcup_{i\in[k]}E(F^o_i\cup F^u_i)$, is called \textbf{update flow network} if the two 
	families $\mathcal{P}^o=\left\{ F_1^o,\dots,F_k^o \right\}$ and 
	$\mathcal{P}^u=\left\{ F_1^u,\dots,F_k^u \right\}$ are valid.
	For an illustration, recall the initial network in Figure \ref{fig:ex2}: 
	The old flow is presented as the directed path made of solid edges 
	and the new one is represented by the dashed edges.
	
	\smallskip
	
	Given an update flow network 
	$G=(V,E,\mathcal{P},s,t,c)$, an \textbf{update} is a pair
	$\mu=(v,P)\in V\times\mathcal{P}$. An update $(v,P)$ with $P=(F^o,F^u)$ is \emph{resolved} 
	by 
	deactivating all outgoing edges of 
	$F^o$ incident to $v$ and activating all of its outgoing edges of $F^u$.
	Note that at all times, 
	there is at most one outgoing and at most one incoming edge, 
	for any flow at a 
	given vertex.
	So the deactivated edges of $F^o$ can no longer be used by the flow pair $P$
	(but now the newly 
	activated edges of $F^u$ can).
	
	For any set of updates $U\subset V\times\mathcal{P}$ and any flow pair 
	$P=(F^o,F^u)\in\mathcal{P}$, $G(P,U)$ is the update flow network consisting exactly of the 
	vertices $V(F^o)\cup 
	V(F^u)$ and the edges of $P$ that are active after resolving all updates in $U$.
	
	%
	
	As an illustration, after the second update in Figure \ref{fig:ex2}, 
	one of the original solid edges is still not deactivated. However, 
	already two of the new edges have become solid (i.e., active). 
	So in the picture of the second update, the set $U=\left\{ (v_1,P),(v_2,P) \right\}$ has been 
	resolved.
	
	
	We are now able to determine, for a given set of updates, which edges
	we can and which edges we cannot use for our routing. In the end, we
	want to describe a process of reconfiguration 
	steps, starting from the \emph{initial state}, in which no 
	update has been resolved, and finishing in a state where the only active edges are
	exactly those of the new flows, of every update flow pair.
	
	The flow pair $P$ is called \textbf{transient} for some set of updates $U\subseteq 
	V\times\mathcal{P}$, if $G(P,U)$ contains a unique valid $(s,t)$-flow $T_{P,U}$.
	
	If there is a family $\mathcal{P}=\left\{ P_1,\dots P_k \right\}$ of update flow pairs with 
	demands $d_1,\dots,d_k$ respectively, we call $\mathcal{P}$ a 
	\textbf{transient family} for a set of updates $U\subseteq
	V\times\mathcal{P}$, if and only if every $P\in\mathcal{P}$ is transient for $U$.
	The family of transient flows after all updates in $U$ are resolved is denoted by 
	$\mathcal{T}_{\mathcal{P},\mathcal{U}}=\left\{ T_{P_1,U},\dots,T_{P_k,U}\right\}$. 
	
	We again refer to Figure~\ref{fig:ex2}. 
	In each of the different states, the transient flow is depicted as the light blue 
	line connecting $s$ to $t$ and covering only solid (i.e., active) edges.

	An \textbf{update sequence} $(\updatesequence_i)_{i\in[\left| V\times\mathcal{P} \right|]}$ is an 
	ordering 
	of $V\times\mathcal{P}$. We denote the set of updates that is resolved after step $i$ 
	by $U_i=\bigcup_{j=1}^i\updatesequence_i$, for all $i\in[\left| V\times\mathcal{P} \right|]$.
	
	\begin{definition}[Consistency Rule]\label{def:consistencyrule} 
		Let $\updatesequence$ be an update sequence.
		We require that for any $i\in[\left| V\times\mathcal{P} \right|]$,
		there is a family of transient flow pairs $\mathcal{T}_{\mathcal{P},\mathcal{U}_i}$.
	\end{definition}
	
	To ease the notation, we will denote an update sequence $(\updatesequence)_{i\in[\left| 
		V\times\mathcal{P} \right|]}$ simply by $\updatesequence$ and for any update $(u,P)$ we 
		write 
	$\updatesequence(u,P)$ for the the position $i$ of $(u,P)$ within $\updatesequence$.
	An update sequence  is \textbf{valid}, if every set $U_i$, $i\in[\left|  V\times\mathcal{P}\right|]$, 
	obeys 
	the consistency rule.
	
	\smallskip
	
	
	
	We note that this consistency rule models and 
	consolidates the fundamental  
	properties usually studied in the literature, such as  
	congestion-freedom~\cite{roger-infocom} and 
	loop-freedom~\cite{ludwig2015scheduling}. 

	Note that we do not forbid edges $e\in E(F^o_i\cap F^u_i)$ and 
	we never activate or deactivate such an edge. Starting with an initial update 
	flow network, these edges will be active and remain so until all updates are resolved. 
	Hence there are vertices $v\in V$ with either no outgoing edge for a given flow pair $F$ at all;
	or with an outgoing edge which however is used 
	by both the old and the update flow of $F$.
	Such updates do not have any impact on the actual problem since they
	never affect a transient flow. Hence they can always be scheduled in
	the first round, and thus w.l.o.g.~we ignore them in the
	following. 
	
	\begin{definition}[\textsc{$k$-Network Flow Update Problem}]
		Given an update flow network $G$ with $k$ update flow pairs, 
		is there a feasible update sequence $\updatesequence$? 
	\end{definition}
	

	\section{NP-Hardness of 2-Flow Update in General
		Graphs}\label{2hard}
	
	It is easy to see that for an update flow network with a single
	flow pair, feasibility is always guaranteed. However, it turns out that for two flows,
	the problem becomes hard in general.
	
	\begin{theorem}\label[theorem]{thm:hardness}
		Deciding whether a feasible network update schedule exists is NP-hard already for $k=2$ 
		flows.
	\end{theorem}
	
	The proof is by reduction from 3-SAT. In what follows 
	let $C$ be any 3-SAT formula with $n$ variables and $m$ clauses.
	We will denote the variables as $X_1, \dots, X_{n}$ and the clauses 
	as $C_1, \dots, C_{m}$.
	The resulting update flow network will be denoted as $G(C)$. 
	Furthermore, we will assume that the variables are ordered by their 
	indices and their appearance in each clause respects this order.
	
	We will create $2$ update flow pairs, a blue one $B=(B^o,B^u)$ 
	and a red one $R=(R^o,R^u)$, both of demand $1$. The pair $B$ will contain
	gadgets corresponding to the variables. The order in which the edges 
	of each of those gadgets are updated will correspond to 
	assigning a value to the variable. The pair $R$ on the other hand will 
	contain gadgets representing the clauses: they will have edges that are 
	``blocked'' by the variable edges of $B$. Therefore, 
	we will need to update $B$ to enable the updates of $R$.
	
	We proceed by giving a precise construction of the 
	update flow network $G(C)$. 
	In the following, the capacities of all edges will be $1$. 
	Since we are working with just two flows and each of those 
	flows contains many gadgets, we give the construction of the 
	two update flow pairs in terms of their gadgets.
	
	\begin{enumerate}
		\item \textbf{Clause Gadgets:} For every $i\in[m]$, we introduce eight vertices 
		$u^{i}_1,u^{i}_2,\dots,u^{i}_8$ corresponding to the clause $C_i$. The edges $(u^i_j,u^i_{j+1})$ 
		with $j\in[7]$ are added to $R^o$ while the edges $(u^i_{j'},u^i_{j'+5})$ for $j'\in\left\{ 1,2,3 
		\right\}$ and $(u^i_{j'},u^i_{j'-4})$ for $j'\in\left\{ 6,7 \right\}$ are added to $R^u$.
		
		\item \textbf{Variable Gadgets:} For every $j\in[n]$, we introduce four
		vertices: $v_1^j,\dots,v_4^j$. Let $P_j=\left\{
		p_1^j,\dots,p_{k_j}^j \right\}$ denote the set of indices of the
		clauses containing the literal $x_j$ and $\overline{P}_j=\left\{
		\overline{p}^j_1,\dots,\overline{p}^j_{k'_j} \right\}$ the set of
		indices of the clauses containing the literal
		$\overline{x}_j$. Furthermore, let $\pi(i,j)$ denote the position of
		$x_j$ in the clause $C_i$, $i\in P_j$. Similarly,
		$\overline{\pi}(i',j)$ denotes the position of $\overline{x_j}$ in $C_{i'}$ where 
		$i'\in\overline{P}_j$.
		
		To $B^o$ we now add the following edges for every $j\in[n]$:
		\begin{enumerate}[i)]	
			\item $(u^i_{\pi(i,j)}, u^i_{\pi(i,j)+5})$, for $i\in P_j$ 
			(these edges are shared with $R^u$),
			\item $(u^i_{\pi(i,j)+5}, u^i_{\pi(i+1,j)})$, for $i\in P_j, i\neq p_{k_j}^j$,
			\item $(v^j_1, u^{p_1^j}_{\pi(p_1^j,j)})$ and $(u^{p_{k_j}^j}_{\pi(p_{k_j}^j,j)+5}, v^j_2)$,
			\item $(u^{i}_{\overline{\pi}(i,j)}, u^{i}_{\overline{\pi}(i,j)+5})$, for $i\in\overline{P}_j$,
			\item $(u^{\overline{p}^j_i}_{\overline{\pi}(\overline{p}^j_i,j)+5}, 
			u^{\overline{p}^j_{i+1}}_{\overline{\pi}(\overline{p}^j_{i+1},j)})$, for 
			$i\in[\sizeof{\overline{P}_j}-1]$,
			\item $(v^j_3, u^{\overline{p}^j_1}_{\overline{\pi}(p^j_1,j)})$ and 
			$(u^{\overline{p}^j_{\bar{k}_j}}_{\overline{\pi}(\overline{p}^j_{\bar{k}_j},j)+5}, v^j_4)$, and
			\item $(v^j_2, v^j_3)$.
		\end{enumerate}
		On the other hand, $B^u$ will contain the edges $(v_1^j,v_3^j)$, 
		$(v_3^j,v_2^j)$ and $(v_2^j,v_4^j)$.
		
		\item[] \begin{figure}[h]
			\begin{center}
				\begin{tikzpicture}[scale=0.62]
				\tikzset{>=latex} 
				
				\node (center) [] {};
				
				\node (a1) [position=180:3mm from center] {};
				\node (a2) [position=0:3mm from center] {};
				
				\foreach\i in {1,...,8} {
					\pgfmathtruncatemacro{\iPos}{9+(8-\i)*9};
					\node (u-\i) [position=180:\iPos mm from a1,circle,draw,scale=0.6,thick] {};
					\node (u-\i-label) [position=270:8mm from u-\i] {$u^i_{\i}$};		
				}
				
				\draw[red, ->,thick] (u-1) -- (u-2);
				\draw[red, ->,thick] (u-2) -- (u-3);
				\draw[red, ->,thick] (u-3) -- (u-4);
				\draw[red, ->,thick] (u-4) -- (u-5);
				\draw[red, ->,thick] (u-5) -- (u-6);
				\draw[red, ->,thick] (u-6) -- (u-7);
				\draw[red, ->,thick] (u-7) -- (u-8);
				
				\draw[red, dashed, ->,thick] (u-1) to [bend left] (u-6);
				\draw[red, dashed, ->,thick] (u-2) to [bend left] (u-7);
				\draw[red, dashed, ->,thick] (u-3) to [bend left] (u-8);
				\draw[red, dashed, ->,thick] (u-6) to [bend left=20] (u-2);
				\draw[red, dashed, ->,thick] (u-7) to [bend left=20] (u-3);
				
				\foreach\i in {1,...,2} {
					\foreach\j in {1,...,2} {
						\pgfmathtruncatemacro{\iPoso}{(\j-1)*45+(\i-1)*57};
						\pgfmathtruncatemacro{\ind}{\j+2*(\i-1)};
						\node (w-\i-\j) [position=0:\iPoso mm from a2,circle,draw,scale=0.8,thick] {};
						\node (w-\i-\j-label) [position=270:6mm from w-\i-\j] {$v_{\ind}^j$};	
					}
					\foreach\k in {1,...,4} {
						\pgfmathtruncatemacro{\iPos}{(\k)*9+(\i-1)*57};
						\node (v-\i-\k) [position=0:\iPos mm from a2,circle,draw,scale=0.6,thick] {};		
					}	
				}
				
				\node (v-1-1-label) [position=270:8mm from v-1-1] {$u_1^{p^j_1}$};
				\node (v-1-2-label) [position=270:8mm from v-1-2] {$u_6^{p^j_1}$};
				\node (v-1-3-label) [position=270:8mm from v-1-3] {$u_1^{p^j_{k_j}}$};
				\node (v-1-4-label) [position=270:8mm from v-1-4] {$u_6^{p^j_{k_j}}$};
				\node (v-2-1-label) [position=270:8mm from v-2-1] {$u_1^{\overline{p}^j_1}$};
				\node (v-2-2-label) [position=270:8mm from v-2-2] {$u_6^{\overline{p}^j_1}$};
				\node (v-2-3-label) [position=270:8mm from v-2-3] {$u_1^{\overline{p}^j_{\bar{k}_j}}$};
				\node (v-2-4-label) [position=270:8mm from v-2-4] {$u_6^{\overline{p}^j_{\bar{k}_j}}$};

				\draw[blue, ->,thick] (w-1-1) -- (v-1-1);
				\draw[blue, ->,thick] (v-1-1) -- (v-1-2);
				\draw[blue, line width = .8pt, dash pattern=on .13pt off 3.3pt, dash phase=2pt, line 
				cap=round] (v-1-2) -- (v-1-3);
				\draw[blue, ->,thick] (v-1-3) -- (v-1-4);
				\draw[blue, ->,thick] (v-1-4) -- (w-1-2);
				
				\draw[blue, ->,thick] (w-1-2) -- (w-2-1);
				\draw[blue, ->,thick] (w-2-1) -- (v-2-1);
				\draw[blue, ->,thick] (v-2-1) -- (v-2-2);
				\draw[blue, line width = .8pt, dash pattern=on .13pt off 3.3pt, dash phase=2pt, line 
				cap=round] (v-2-2) -- (v-2-3);
				\draw[blue, ->,thick] (v-2-3) -- (v-2-4);
				\draw[blue, ->,thick] (v-2-4) -- (w-2-2);
				
				\draw[blue, dashed, ->,thick] (w-1-1) to [bend left] (w-2-1);
				\draw[blue, dashed, ->,thick] (w-2-1) to [bend left] (w-1-2);
				\draw[blue, dashed, ->,thick] (w-1-2) to [bend left] (w-2-2);
				
				\node (v4j-1) [draw,circle,thick,position=110:12mm from w-1-1,scale=0.5] {};
				\node (labelv4j-1) [position=20:7mm from v4j-1] {$v_4^{j-1}$};
				\node (v4j-11) [position=180:0.5mm from v4j-1,scale=0.55] {};
				\node (v4j-12) [position=0:0.5mm from v4j-1,scale=0.55] {};
				\node (w111) [position=180:0.5mm from w-1-1,scale=0.9] {};
				\node (w112) [position=0:0.5mm from w-1-1,scale=0.9] {};
				\draw [blue,->,thick] (v4j-11) to (w111);
				\draw [blue,->,thick,dashed] (v4j-12) to (w112);
				
				\node (v1j+1) [draw,circle,thick,position=70:12mm from w-2-2,scale=0.5] {};
				\node (labelv1j+1) [position=160:7mm from v1j+1] {$v_1^{j+1}$};
				\node (v1j+11) [position=180:0.5mm from v1j+1,scale=0.55] {};
				\node (v1j+12) [position=0:0.5mm from v1j+1,scale=0.55] {};
				\node (w221) [position=180:0.5mm from w-2-2,scale=0.9] {};
				\node (w222) [position=0:0.5mm from w-2-2,scale=0.9] {};
				\draw [blue,->,thick] (w221) to (v1j+11);
				\draw [blue,->,thick,dashed] (w222) to (v1j+12);
				
				\node (u8i-1) [draw,circle,thick,position=110:12mm from u-1,scale=0.5] {};
				\node (labelu8i-1) [position=20:7mm from u8i-1] {$u_8^{i-1}$};
				\node (u8i-11) [position=180:0.5mm from u8i-1,scale=0.55] {};
				\node (u8i-12) [position=0:0.5mm from u8i-1,scale=0.55] {};
				\node (u11) [position=180:0.5mm from u-1,scale=0.9] {};
				\node (u12) [position=0:0.5mm from u-1,scale=0.9] {};
				\draw [red,->,thick] (u8i-11) to (u11);
				\draw [red,->,thick,dashed] (u8i-12) to (u12);
				
				\node (u1i+1) [draw,circle,thick,position=70:12mm from u-8,scale=0.5] {};
				\node (labelu1i+1) [position=160:7mm from u1i+1] {$u_1^{i+1}$};
				\node (u1i+11) [position=180:0.5mm from u1i+1,scale=0.55] {};
				\node (u1i+12) [position=0:0.5mm from u1i+1,scale=0.55] {};
				\node (u81) [position=180:0.5mm from u-8,scale=0.9] {};
				\node (u82) [position=0:0.5mm from u-8,scale=0.9] {};
				\draw [red,->,thick] (u81) to (u1i+11);
				\draw [red,->,thick,dashed] (u82) to (u1i+12);
				
				\end{tikzpicture}
				\caption{\emph{Examples:} A clause gadget is shown in red,
					the $R^o$ edges are depicted as a solid line, and
					the dashed lines belong to $R^u$. The variable gadget 
					is shown in blue. Again, solid lines indicate the old flow
					and 
					dashed lines the update flow.}
				\label{fig:gadgets}
			\end{center}
		\end{figure}
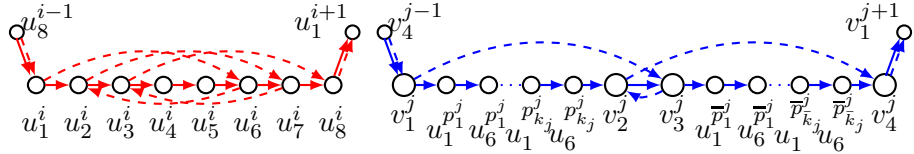
		
		\item\textbf{Blocking Edges:} The goal is to block the updates 
		$(v_3^j,B)$ for every $j\in[n]$ until all clauses are satisfied. 
		To do this, we introduce $4$ additional vertices $w_1$, $w_2$, $z_1$ and $z_2$. 
		Then for $R^o$, we introduce the following edges:
		\begin{enumerate}[i)]
			\item $(v_3^j,v_2^j)$ for $j\in[n]$,	
			\item $(v_2^j,v_3^{j+1})$ for $j\in[n-1]$, and
			\item $(z_1,v_3^j)$ and $(v_2^n,z_2)$,
		\end{enumerate}	
		while $R^u$ contains the edges $(z_1)$, $(w_1,w_2)$ and $(w_2,z_2)$.
		
		In a similar fashion, $B^o$ contains the edge $(w_1,w_2)$.
		For $B^u$, 
		we introduce the following edges:
		\begin{enumerate}[i)]
			\item $(u_4^i,u_5^i)$ for $i\in[m]$,
			\item $(u_5^i,u_4^{i+1})$ for $i\in[m-1]$, and
			\item $(w_1,u_4^1)$ and $(u_5^m,w_2)$.
		\end{enumerate}
		
		\item[]
		
		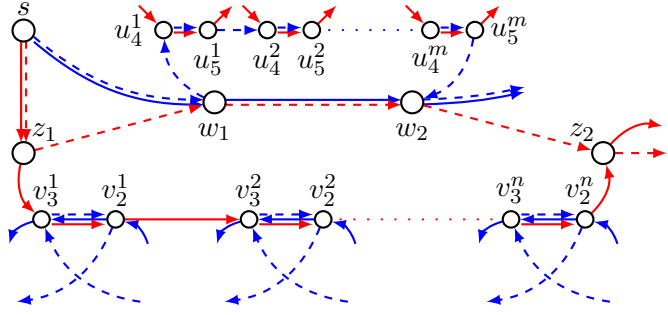
\begin{figure}[h]
			\begin{center}
				\begin{tikzpicture}[scale=0.65]
				\tikzset{>=latex} 
				
				\node (z1) [circle,draw,scale=0.8,thick] {};
				\node (w1) [circle,draw,scale=0.8,position=15:40mm from z1,thick] {};
				\node (w2) [circle,draw,scale=0.8,position=0:40mm from w1,thick] {};
				\node (w1b) [position=90:0.5mm from w1] {};
				\node (w2b) [position=90:0.5mm from w2] {};
				\node (w1r) [position=270:0.5mm from w1] {};
				\node (w2r) [position=270:0.5mm from w2] {};
				\node (z2) [circle,draw,scale=0.8,position=345:40mm from w2,thick] {};
				
				\node (s) [circle,draw,scale=0.8,thick,position=90:25mm from z1] {};
				\node (s1) [position=0:0.5mm from s] {};
				\node (s2) [position=180:0.5mm from s] {};
				\node (z11) [position=0:0.5mm from z1] {};
				\node (z12) [position=180:0.5mm from z1] {};
				\node (so) [position=90:0.5mm from s] {};
				\node (su) [position=270:0.5mm from s] {};
				\node (w1o) [position=90:0.5mm from w1] {};
				\node (w1u) [position=270:0.5mm from w1] {};
				\node (labels) [position=90:4.5mm from s] {$s$};
				
				\node (gz1) [position=180:15mm from z1] {};
				\node (gw1) [position=175:15mm from w1] {};
				\node (gw2) [position=7:25mm from w2] {};
				\node (gw2o) [position=90:0.5mm from gw2] {};
				\node (gw2u) [position=270:0.5mm from gw2] {};
				\node (gz2) [position=0:15mm from z2] {};
				\node (ggz1) [position=160:15mm from z1] {};
				\node (ggz2) [position=20:15mm from z2] {};
				
				\node (u41) [circle,draw,scale=0.6,position=125:18mm from w1,thick] {};
				\node (u51) [circle,draw,scale=0.6,position=0:9mm from u41,thick] {};
				
				\node (u41b) [position=90:0.5mm from u41] {};
				\node (u51b) [position=90:0.5mm from u51] {};
				\node (u41r) [position=270:0.5mm from u41] {};
				\node (u51r) [position=270:0.5mm from u51] {};
				
				\node (u42) [circle,draw,scale=0.6,position=0:12mm from u51,thick] {};
				\node (u52) [circle,draw,scale=0.6,position=0:9mm from u42,thick] {};
				
				\node (u42b) [position=90:0.5mm from u42] {};
				\node (u52b) [position=90:0.5mm from u52] {};
				\node (u42r) [position=270:0.5mm from u42] {};
				\node (u52r) [position=270:0.5mm from u52] {};
				
				\node (u4m) [circle,draw,scale=0.6,position=0:24mm from u52,thick] {};
				\node (u5m) [circle,draw,scale=0.6,position=0:9mm from u4m,thick] {};
				
				\node (u4mb) [position=90:0.5mm from u4m] {};
				\node (u5mb) [position=90:0.5mm from u5m] {};
				\node (u4mr) [position=270:0.5mm from u4m] {};
				\node (u5mr) [position=270:0.5mm from u5m] {};
				
				\node (gu41) [position=135:10mm from u41] {};
				\node (gu51) [position=45:10mm from u51] {};
				
				\node (gu42) [position=135:10mm from u42] {};
				\node (gu52) [position=45:10mm from u52] {};
				
				\node (gu4m) [position=135:10mm from u4m] {};
				\node (gu5m) [position=45:10mm from u5m] {};
				
				\node (v31) [draw,circle,scale=0.6,position=285:14mm from z1,thick] {};
				\node (v21) [draw,circle,scale=0.6,position=0:15mm from v31,thick] {};
				\node (gv31) [position=225:10mm from v31] {};
				\node (ggv31) [position=270:10mm from gv31] {};
				\node (gv21) [position=315:10mm from v21] {};
				\node (ggv21) [position=270:10mm from gv21] {};
				\node (v31b) [position=90:1mm from v31] {};
				\node (v31r) [position=270:1mm from v31] {};
				\node (v21b) [position=90:1mm from v21] {};
				\node (v21r) [position=270:1mm from v21] {};
				
				\node (v32) [draw,circle,scale=0.6,position=0:27mm from v21,thick] {};
				\node (v22) [draw,circle,scale=0.6,position=0:15mm from v32,thick] {};
				\node (gv32) [position=225:10mm from v32] {};
				\node (ggv32) [position=270:10mm from gv32] {};
				\node (gv22) [position=315:10mm from v22] {};
				\node (ggv22) [position=270:10mm from gv22] {};
				\node (v32b) [position=90:1mm from v32] {};
				\node (v32r) [position=270:1mm from v32] {};
				\node (v22b) [position=90:1mm from v22] {};
				\node (v22r) [position=270:1mm from v22] {};
				
				\node (v3n) [draw,circle,scale=0.6,position=0:38mm from v22,thick] {};
				\node (v2n) [draw,circle,scale=0.6,position=0:15mm from v3n,thick] {};
				\node (gv3n) [position=225:10mm from v3n] {};
				\node (ggv3n) [position=270:10mm from gv3n] {};
				\node (gv2n) [position=315:10mm from v2n] {};
				\node (ggv2n) [position=270:10mm from gv2n] {};
				\node (v3nb) [position=90:1mm from v3n] {};
				\node (v3nr) [position=270:1mm from v3n] {};
				\node (v2nb) [position=90:1mm from v2n] {};
				\node (v2nr) [position=270:1mm from v2n] {};
				
				\node (labelz1) [position=45:6mm from z1] {$z_1$};
				\node (labelw1) [position=270:6mm from w1] {$w_1$};
				\node (labelw2) [position=270:6mm from w2] {$w_2$};
				\node (labelz2) [position=135:6mm from z2] {$z_2$};
				
				\node (labelz1) [position=180:7mm from u41] {$u_4^1$};
				\node (labelz1) [position=270:6mm from u51] {$u_5^1$};
				\node (labelz1) [position=270:6mm from u42] {$u_4^2$};
				\node (labelz1) [position=270:6mm from u52] {$u_5^2$};
				\node (labelz1) [position=270:6mm from u4m] {$u_4^m$};
				\node (labelz1) [position=0:7mm from u5m] {$u_5^m$};
				
				\node (labelz1) [position=80:6mm from v31] {$v_3^1$};
				\node (labelz1) [position=90:6mm from v21] {$v_2^1$};
				\node (labelz1) [position=90:6mm from v32] {$v_3^2$};
				\node (labelz1) [position=90:6mm from v22] {$v_2^2$};
				\node (labelz1) [position=90:6mm from v3n] {$v_3^n$};
				\node (labelz1) [position=100:6mm from v2n] {$v_2^n$};
				
				\begin{pgfonlayer}{bg}		
				
				\draw[blue,->,thick,bend right=20] (su) to (w1u);
				\draw[blue,->,thick,bend right=20,dashed] (so) to (w1o);
				\draw[blue,->,thick] (w1b) to (w2b);
				\draw[blue,->,thick,bend right=5] (w2r) to (gw2u);
				\draw[blue,->,thick,dashed,bend right=5] (w2b) to (gw2o);
				
				\draw[red,dashed,->,thick] (s1) to (z11);
				\draw[red,dashed,->,thick] (z1) to (w1);
				\draw[red,dashed,->,thick] (w1r) to (w2r);
				\draw[red,dashed,->,thick] (w2) to (z2);
				\draw[red,dashed,->,thick] (z2) to (gz2);
				
				\draw[red,->,thick] (gu41) to (u41);
				\draw[red,->,thick] (u41r) to (u51r);
				\draw[red,->,thick] (u51) to (gu51);
				
				\draw[red,->,thick] (gu42) to (u42);
				\draw[red,->,thick] (u42r) to (u52r);
				\draw[red,->,thick] (u52) to (gu52);
				
				\draw[red,->,thick] (gu4m) to (u4m);
				\draw[red,->,thick] (u4mr) to (u5mr);
				\draw[red,->,thick] (u5m) to (gu5m);
				
				\draw[blue, dashed, ->,bend left,thick]  (w1) to  (u41);
				\draw[blue, dashed, ->,thick]  (u41b) to (u51b);
				\draw[blue, dashed, ->,thick]  (u51) to (u42);
				\draw[blue, dashed, ->,thick]  (u42b) to (u52b);
				\draw[blue, line width = .8pt, dash pattern=on .13pt off 5pt, dash phase=2pt, line 
				cap=round]  (u52) to (u4m);
				\draw[blue, dashed, ->,thick]  (u4mb) to (u5mb);
				\draw[blue, dashed, ->,bend left,thick]  (u5m) to (w2);	
				
				\draw [blue,->,thick,bend right] (gv21) to (v21);
				\draw [blue,->,thick] (v21) to (v31);
				\draw [blue,->,thick,bend right] (v31) to (gv31);
				\draw [blue,dashed,->,thick,bend left] (ggv21) to (v31);
				\draw [blue,dashed,->,thick] (v31b) to (v21b);
				\draw [blue,dashed,->,thick,bend left] (v21) to (ggv31);
				\draw [red,->,thick] (v31r) to (v21r);
				
				\draw [blue,->,thick,bend right] (gv22) to (v22);
				\draw [blue,->,thick] (v22) to (v32);
				\draw [blue,->,thick,bend right] (v32) to (gv32);
				\draw [blue,dashed,->,thick,bend left] (ggv22) to (v32);
				\draw [blue,dashed,->,thick] (v32b) to (v22b);
				\draw [blue,dashed,->,thick,bend left] (v22) to (ggv32);
				\draw [red,->,thick] (v32r) to (v22r);
				
				\draw [blue,->,thick,bend right] (gv2n) to (v2n);
				\draw [blue,->,thick] (v2n) to (v3n);
				\draw [blue,->,thick,bend right] (v3n) to (gv3n);
				\draw [blue,dashed,->,thick,bend left] (ggv2n) to (v3n);
				\draw [blue,dashed,->,thick] (v3nb) to (v2nb);
				\draw [blue,dashed,->,thick,bend left] (v2n) to (ggv3n);
				\draw [red,->,thick] (v3nr) to (v2nr);
				
				\draw [red,->,thick,bend right] (z1) to (v31);
				\draw [red,->,thick,bend right] (v2n) to (z2);
				\draw [red,->,thick] (v21) to (v32);
				\draw [red, line width = .8pt, dash pattern=on .13pt off 5pt, dash phase=2pt, line 
				cap=round]  (v22) to (v3n);
				\draw [red,->,thick] (s2) to (z12);
				\draw [red,->,thick,bend left] (z2) to (ggz2);
				
				\end{pgfonlayer}
				
				\end{tikzpicture}
				\caption{The gadget for blocking the update  $(v_3^j,B)$ for all $j\in[n]$. Again dashed 
				edges correspond to the update flows and solid ones to the old flows.}
				\label{fig:blocking_gadget}
			\end{center}
		\end{figure}
		
		\item \textbf{Source and Terminal.} Finally, to complete the graph, 
		we introduce a source $s$ and a terminal $t$.
		
		For both, $R^o$ and $R^u$ we introduce the following edges:
		\begin{enumerate}[i)]
			\item $(s,z_1)$ and $(z_2,u_1^1)$,
			\item $(u_8^i,u_1^{i+1})$ for $i\in[m-1]$, and
			\item $(u_8^m,t)$.	
		\end{enumerate}
		And for $B^o$ and $B^u$ we complete the flows with the following edges:
		\begin{enumerate}[i)]
			\item $(s,w_1)$ and $(w_2,v_1^1)$,
			\item $(v_4^j,v_1^{j+1})$ for $j\in[n-1]$, and
			\item $(v_4^n,t)$.
		\end{enumerate}
		
	\end{enumerate}

	\begin{lemma} Given any valid update sequence $\updatesequence$ 
		for the above constructed update flow network $G(C)$, 
		the following conditions hold for every $r<\updatesequence(w_1,B)$. 
		\begin{enumerate}
			\item $r<\updatesequence(z_1,R)$ \label{con:z1notupdated}
			\item For any $j\in[n]$, $v^j_1$ is a vertex of the transient network flow $T_{B,U_r}$ and 
			$r<\updatesequence(v^j_3,B)$.
			\label{con:reachabilityvx1}
			\item Let $j\in[n]$ and $P_j$ and $\overline{P}_j$ be the 
			index sets of the clauses containing the corresponding literals 
			$x_j$ and $\overline{x}_j$. Then $T_{B,U_r}$ contains all edges of the form 
			$(u^i_{\pi(i,j)},u^i_{\pi(i,j)+5})$ for $i\in P_i$, or all the edges 
			$(u^i_{\overline{\pi}(i,j)},u^i_{\overline{\pi}(i,j)+5})$ for 
			$i\in\overline{P}_j$ (or both).
			\label{con:varuseclause}
			\item The vertex $z_1$ and the $u_1^i$, for all $i\in[m]$,  are contained in $T_{R,U_r}$.
			\label{con:clausereachability}
		\end{enumerate}
		\label{lem:2flowhardaux}
	\end{lemma}
	\begin{proof}
		\begin{enumerate}
			\item Suppose $\updatesequence(z_1,R)\leq r$, then there is a step $r'\geq r$ such that 
			$(w_1,B)$ 
			is not in $U_{r'}$, but $(z_1,R)$ is. If $\updatesequence(w_1,R)\leq r'$,	$T_{R,U_{r'}}$ and 
			$T_{B,U_{r'}}$ pass through $(w_1,w_2)$ violating the capacity of $1$, 
			otherwise there is no path $T_{R,U}$ in $G(R,U)$.
			
			\item The first assertion is trivially true, since the edges $(w_2,v_1^1)$ and 
			$(v_4^j,v_1^{j+1})$ for $j\in[n-1]$ belong to both $B^o$ and $B^u$, 
			hence $T_{B,U_r}$ has to always contain these edges. From 
			Property~\ref{con:z1notupdated} 
			we know, that $T_{R,U_r}$ has to contain the $z_1$-$z_2$-subpath of $R^o$ and thus 
			$T_{R,U_r}$ 
			fills the capacity of the edges $(v_3^j,v_2^j)$ for all $j\in[n]$: hence resolving $(v_3^j,B)$ is 
			impossible for all $j\in[n]$.
			
			\item Let $j\in[n]$. By Property~\ref{con:reachabilityvx1}, 
			$v_1^j$ is contained in $T_{R,U_r}$, but $\updatesequence(v_3^j,B)>r$. Hence, if 
			$\updatesequence(v_1^j,B)\leq r$, then $T_{B,U_r}$ traverses directly from $v_1^j$ to 
			$v_3^j$ 
			and then follows along $B^o$ to $v_4$. Otherwise it follows 
			along $B^o$ from $v_1^j$ to $v_3^j$. In both cases we are done.
			
			\item This is again trivially true, since the edges 
			$(s,z_1)$ and $(u_8^i,u_1^{i+1})$ for $i\in[m-1]$ are 
			contained in both $R^o$ and $R^u$: thus they always 
			have to be part of $T_{R,U_r}$.
		\end{enumerate}
	\end{proof}
	
	\begin{proof} [Proof of~\Cref{thm:hardness}]
		Now we are ready to finish the proof of~\Cref{thm:hardness}. 
		First we will show that if $C$ is satisfiable, then there is a feasible 
		order of updates for $G(C)$.
		Let $\sigma$ be an assignment satisfying $C$. Then the update order for $G(C)$ is as 
		follows. For 
		each item $i$ we define $r_i^f$ to be the position of the first update defined by $i$ and 
		$r_i^l$ to be 
		the position of its last update:
		\begin{enumerate}
			\item For each $j\in[n]$, if $\sigma(X_j)=1$ 
			then update $v^j_1$. Otherwise update $v^j_2$.
			\item For each $i\in[m]$, at least one of edges 
			$(u^i_1, u^i_6),(u^i_2, u^i_7),(u^i_3, u^i_8)$ is no longer 
			used by $T_{B,U_{r_2^f-1}}$. Therefore the corresponding update of $R$ can 
			be resolved (this follows from $\sigma$ being a satisfying assignment).
			\item For each $i\in[m]$, $(u^i_4,u^i_5)$ is no longer used by $T_{R,U_{r_3^f-1}}$. 
			Therefore we can resolve to blue updates along the $w_1$-$w_2$-subpath 
			of $B^u$ excluding $(w_1,B)$. 
			\item Resolve $(w_1,B)$.
			\item Resolve $(w_1,R)$ and $(w_2,R)$. (Note that now all conflicts between $B$ and $R$ 
			have been resolved and we can finish the updates. We will now leave the state described 
			in~\Cref{lem:2flowhardaux}.)
			\item Resolve $(z_1,R)$.
			\item For each $j\in[n]$, $v_k^j$ has already been updated for exactly one $k\in\left\{ 1,2 
			\right\}$. If $k=1$, resolve all updates of $B$ along the 
			$u_1^{p_1^j}$-$u_6^{p^j_{k_j}}$-subpath of $B^o$ together with $(v_2^j,B)$. Otherwise 
			resolve $(v_3^j,B)$ together with all updates of $B$ along the 
			$u_1^{\overline{p}_1^j}$-$u_6^{\overline{p}^j_{\overline{k}_j}}$-subpath of $B^o$.
			\item Resolve the remaining updates of $B$. 
			\item Resolve all updates of $R$ along the $v_3^1$-$v_2^n$-subpath 
			of $R^o$ and for each $i\in[m]$ resolve $(u^i_1,R)$, $(u^i_2,R)$ and 
			$(u^i_3,R)$.
			\item Resolve the remaining updates of $R$.
		\end{enumerate}
		
		Now let us assume that there is a feasible update sequence 
		$\updatesequence$ for $G(C)$. We will show that $C$ is satisfiable 
		by constructing an assignment $\sigma$.
		
		Let us consider the steps $r<\min\left\{ \updatesequence(w_1,R),\updatesequence(w_1,B) 
		\right\}$. 
		Then we will use Condition~\ref{con:varuseclause} of~\Cref{lem:2flowhardaux} to assign 
		values to 
		variables in the following way.
		Let $j\in[n]$, if $T_{B,U_r}$ does not use the edges $(u^h_{\pi(h,j)}, u^h_{\pi(h,j)+5})$ for all 
		$h\in 
		P_j$ (or equivalently if $v^j_1$ is updated) we set $\sigma(X_j)\coloneqq1$. Otherwise we set 
		$\sigma(x)\coloneqq0$.
		
		Now we will show that assignment $\sigma$ satisfies $C$.
		First let us notice that because we can resolve $(w_1,B)$, 
		none of edges $(u^i_4,u^i_5)$, for any $i\in[m]$, can be used by 
		$T_{B,U_{\updatesequence(w_1,B)}}$ in $\updatesequence(w_1,B)$. Hence, from 
		Condition~\ref{con:clausereachability} of~\Cref{lem:2flowhardaux}, we know that all 
		vertices $u^i_1$, for any $i\in[n]$, and the vertex $z_1$, 
		are contained in $T_{R,U_{\updatesequence(w_1,B)}}$. 
		
		Let us consider any clause $C_i$, $i\in[m]$. 
		The transient network flow $T_{R,U_r}$ cannot go from 
		$u^{i}_1$ to $u^{i+1}_1$ along $R^o$: 
		this would mean that edge $(u^{i}_4, u^{i}_5)$ cannot be 
		used by $T_{B,U_r}$. Therefore, for at least one of the edges $(u^{i}_1,u^{i}_6)$, 
		$(u^{i}_2,u^{i}_7)$ and $(u^{i}_3,u^{i}_8)$, 
		the corresponding blue update has already been resolved. 
		This implies that there is some variable $X_j$, $j\in[n]$, that appears in $C_i$, 
		such that, in the gadget for $X_j$, $T_{B,U_r}$ skips $u^{i}_h$, for some $h\in\{1,2,3\}$. 
		This vertex is between $v^j_1$ and $v^j_2$, if $C_i$ contains literal $x_j$.
		In that case, we set $\sigma(X_j)\coloneqq1$, so $C_i$ is satisfied. 
		Otherwise $C_i$ contains literal $\bar{x}_j$ and we 
		assign $\sigma(X_j)\coloneqq0$, so $C_i$ is also satisfied.
	\end{proof}

	\section{Rerouting flows in DAGs}
        In this section we consider the flow rerouting problem when
        the underlying flow graph is acyclic. In the remainder of this
        work we will always consider our update flow network to be acyclic.  
	This leads to an important substructure in the flow pairs: the blocks.
	These blocks will play a major role in both the hardness proof and the algorithm.
		
	Let $G=(V,E,\mathcal{P},s,t,c)$ be an acyclic update flow network, i.e., 
	we assume that the graph $(V,E)$ is acyclic. Let $\prec$ be a topological 
	order on the vertices $V=\left\{ v_1,\dots,v_n \right\}$. 
	Let $P_i=(F^o_i,F^u_i)$ be an update flow 
	pair of demand $d$ and let $v_1^i,\dots,v_{\ell_i^o}^i$ be the induced topological 
	order on the vertices of $F^o_i$; analogously,
	let $u_1^i,\dots,v_{\ell^u_i}^i$ 
	be the order on $F^u_i$. Furthermore, let 
	$V(F^o_i)\cap V(F^u_i)=\left\{ z_1^i,\dots,z^i_{k_i} \right\}$ 
	be ordered by $\prec$ as well.
	
	The subgraph of $F_i^o\cup F_i^u$ induced by the set $\left\{ v\in
	V(F_i^o\cup F_i^u) ~|~ z_j^i \prec v \prec z^i_{j+1} \right\}$,
	$j\in[k_i-1]$, is called the $j$th \emph{block} of the update flow
	pair $F_i$, or simply the $j$th \emph{$i$-block}. We 
	will denote this block by $b^i_j$.
	
	For a block $b$, we define $\Start{b}$ to be 
	the \emph{start of the block}, i.e., the smallest vertex
	w.r.t.~$\prec$; similarly, $\End{b}$ is the \emph{end of the block}:
	the largest vertex w.r.t.~$\prec$. 
	
	Let $G=(V,E,\mathcal{P},s,t,c)$ be an update flow network with
	$\mathcal{P}=\left\{ P_1,\dots,P_k \right\}$ and let
	$\blockset$ be the set of its
	blocks. We define a binary relation $<$ between two blocks as follows. 
	For two blocks $b_1,b_2\in \blockset$, where $b_1$ is an $i$-block and $b_2$ a
	$j$-block, $i,j\in[k]$, we say $b_1<b_2$ ($b_1$ \emph{is smaller than}
	$b_2$) if one of the following holds.
	\begin{enumerate}[i]
		\item $\Start{b_1} \prec \Start{b_2}$,
		\item if $\Start{b_1}=\Start{b_2}$ then $b_1<b_2$, if $\End{b_1} \prec \End{b_2}$,
		\item if $\Start{b_1}=\Start{b_2}$ and $\End{b_1}=\End{b_2}$ then $b_1<b_2$, if $i<j$.
	\end{enumerate}
	Let $b$ be an $i$-block and $P_i$ the corresponding update flow pair. 
	For a feasible update sequence $\updatesequence$, we will denote the round 
	$\updatesequence(\Start{b},P_i)$ by $\updatesequence(b)$. We say that $i$-block $b$ is
	\emph{updated}, if all edges in $b\cap F^u_i$ are active and all edges
	in $b\cap F_i^o\setminus F_i^u$ are inactive. 
We will make use of a basic, but important observation on the structure of blocks and how 
	they can 
	be updated.
	This structure is the fundamental idea of the algorithm in the next section since it allows us to 
	consider 
	the update of blocks as a whole instead of updating it vertex by vertex.
	
	\begin{lemma}\label[lemma]{lem:updateblockstart}
		Let $b$ be a block of the flow pair $P=(F^u,F^o)$. Then in
		a feasible update sequence $\updatesequence$, all vertices
		(resp.~their outgoing edges belonging to $P$)
		in $F^u\cap b - \Start{b}$ are
		updated  strictly before $\Start{b}$. Moreover, all vertices in $b-F^u$ are updated strictly
		after $\Start{b}$ is updated.
	\end{lemma}
	
	\begin{proof}
		By $F^u_b$ and $F^o_b$ we denote $F^u\cap b$ and $F^o\cap b$ respectively.
		For the sake of contradiction, let $U=\{v\in V(G)\mid v\in 
		F^u_b-F^o_b-\Start{b},\updatesequence(v,P) 
		> \updatesequence(\Start{b},P)\}$.
		Moreover, let $v$ be the
		vertex of $U$ which is updated the latest
		and $\updatesequence(v,P) = \max_{u\in U}\updatesequence(u,P)$.
		By our condition, the update of $v$ enables 
		a transient flow along
		edges in $F^u_b$. Hence, 
		there now exists an $(s,t)$-flow through $b$ using only update edges.
		
		No vertex in $F_1\coloneqq F^o_b-(F^u_b-\Start{b})$ 
		could have been updated before, or
		simultaneously with $v$: 
		otherwise, between the time $u$ has been updated
		and before the update
		of $v$, there would not exist a transient flow.
		But once we update $v$ in round $r$, there is a transient flow $T_{P,U_r}$ which 
		traverses the vertices in $F^o_b-F^u_b$, and another transient flow $T_{P,U_r}$	traverses
		$v\not\in F_1$: a contradiction. Note that $F_1\neq \emptyset$.
		The other direction is obvious: updating
		any vertex in $(F^o_c\cap b)-F^u_c$ inhibits any transient flow.
	\end{proof}
	
	\begin{lemma}\label[lemma]{lem:updatewholeblock}
		Let $G$ be an update flow network and $\updatesequence$ a valid update sequence for $G$.
		Then there exists a feasible update sequence $\updatesequence'$  which updates every 
		block in 
		consecutive rounds.
	\end{lemma}
	\begin{proof}
		Let $\updatesequence$ be a feasible update sequence with a minimum number of 
		blocks that are not updated in consecutive rounds. 
		Furthermore let $b$ be such a block for the flow pair $P=(F^o,F^u)$.
		Let $r$ be the step in which $\Start{b}$ is updated.
		Then by~\Cref{lem:updateblockstart}, all other vertices of
		$F^u_c\cap b$ have been updated in the previous rounds.
		Moreover, since they do not carry any flow during these rounds, the edges can all be updated 
		in the 
		steps immediately preceding $r$ in any order.
		By our assumption, we can update $\Start{b}$ in round $r$, and hence now this is still 
		possible.
		
		As $\Start{b}$ is updated in step~$r$, the edges of $F^o_c\cap b$ are not used by 
		$T_{P,U_{r+1}}$ 
		and thus we can deactivate all remaining such edges in the steps starting with $r+1$.
		This is a contradiction to the choice of $\updatesequence$, and hence there is always a 
		feasible 
		sequence $\updatesequence'$ satisfying the requirements of the lemma.
	\end{proof}
	
	Note that $G$ is acyclic and every flow pair in $G$ forms a
	single block. Let $\updatesequence$ be a feasible update sequence of $G$. We
	suppose in $\updatesequence$, every block is updated in consecutive
	rounds (\Cref{lem:updatewholeblock}). For a single flow $F$, 
	we write $\updatesequence(F)$ for the round where the last edge of $F$ was updated.

\subsection{Linear Time Algorithm for Constant Number 
		of Flows on DAGs}\label{k-flows}
	
	In the next section we will see that for an arbitrary number of flows,
	the congestion-free flow reconfiguration problem
	is hard, even
	on DAGs. In this section we show that if
	the number of flows is a constant $k$, then a solution 
	can be computed in linear time. More precisely, we
	describe an algorithm to solve the network update
	problem on DAGs in time $2^{O(k\log k)} O(\sizeof{G})$,
	for arbitrary $k$.  In the remainder of this
	section, we assume that every block has at least~$3$ vertices
	(otherwise, postponing such block updates 
	will not affect the solution).
	
	We say a block $b_1$ \emph{touches} a block $b_2$ (denoted by
	$b_1\touches b_2$) if there is a vertex $v\in b_1$ such that
	$\Start{b_2} \prec v \prec \End{b_2}$, or there is a vertex $u\in b_2$ such that
	$\Start{b_1} \prec v \prec \End{b_1}$.
	If $b_1$ does not touch $b_2$, we write $b_1\not\touches b_2$. Clearly, 
	the relation is symmetric, i.e., if $b_1\touches b_2$ then $b_2\touches b_1$.
	
	For some intuition, consider a drawing of $G$ which orders 
	vertices w.r.t.~$\prec$ in a line. Project every edge on that line as well. 
	Then two blocks touch each other if they have a common segment on
	that projection.
	
        \textbf{Algorithm and Proof Sketch}

	Before delving into details, we provide the main ideas behind
	our algorithm. We can think about the update problem 
	on DAGs as follows. Our goal is to compute a feasible
	update order for the (out-)edges of the graph. 
	There are at most $k$ flows to be updated
	for each edge, resulting in $k!$ possible orders 
	and hence a brute force complexity of $O(k!^{\sizeof{G}})$
	for the entire problem. We can reduce this complexity by
	considering blocks instead of edges. 
	
	The update of a given $i$-block $b_i$ might depend on the update of a
	$j$-block sharing at least one edge of $b_i$.
	These dependencies can be represented as a directed graph.
	If this graph does not have any directed cycles, it is rather easy to find a
	feasible update sequence, by iteratively updating sink vertices.
	
	There are several issues here: 
	First of all these dependencies are not straight-forward to define. As
	we will see later, they may lead to representation graphs of
	exponential size.
	In order to control the size we might have to relax our definition of
	dependency, but this might lead to a not necessarily acyclic graph
	which will then need further refinement.
	This refinement is realized by finding a suitable subgraph, which 
	alone is a hard problem in general.
	To overcome the above problems, we proceed as follows.
	
	Let
	$\BTS{b}$ contain all feasible update sequences for 
	the blocks that touch $b$: still a (too) large number, 
	but let us consider them for now. 
	For two distinct blocks $b,b'$, we say that two sequences $s\in
	\BTS{b}, s'\in \BTS{b'}$ are \emph{consistent}, if the order of any common pair of
	blocks is the same in both $s,s'$. It is clear that  if for some block
	$b$, $\BTS{b} = \emptyset$, there is no feasible update sequence
	for $G$: $b$ cannot be updated.
	
	We now consider a graph $H$ whose vertices 
	correspond to elements of $\BTS{b}$, for all $b\in\blockset$. Connect all
	pairs of vertices originating from the same $\BTS{b}$. Connect all pairs
	of vertices if they correspond to inconsistent elements of
	different $\BTS{b}$. If (and only if) we find an independent set of
	size $\sizeof{\blockset}$
	in the resulting graph, the update orders corresponding to those vertices 
	are mutually consistent: we can update the entire network 
	according to those orders. In other words, the update problem
	can be reduced to finding an independent set in the graph $H$. 
	
	However, there are two main issues with this approach. 
	First, $H$ can be very large. A single $\BTS{b}$ can have
	exponentially many elements. Accordingly, we observe that
	we can assume a slightly different perspective on our problem:
	we linearize the lists $\BTS{b}$ and define them sequentially, 
	bounding their size by a function of $k$ (the number of
	flows).
	The second issue is that finding a
	maximum independent set in $H$ is hard.
	The problem is equivalent to finding
	a clique in the complement of $H$, a 
	$\sizeof{\blockset}$-partite graph where every partition has bounded cardinality. 
	We can prove that for an $n$-partite graph where every
	partition has bounded cardinality, finding an $n$-clique is
	NP-complete. So, in order to solve the problem, we either should
	reduce the number of partitions in $H$ (but we cannot) or modify $H$
	to some other graph, further reducing the complexity of the problem. We do the
	latter by trimming $H$ and removing some extra edges, turning
	the graph into a very simple one: 
	a graph
	of \emph{bounded path width}. 
	Then, by standard dynamic programming, 
	we find the
	independent set of size $\sizeof{\blockset}$ in the trimmed version of $H$: this
	independent set matches the independent set $I$ of size $\sizeof{\blockset}$ in $H$ (if it
	exists). At the end, reconstructing a correct update order sequence from
	$I$ needs some effort. As we have reduced the size of $\BTS{b}$ and
	while not all possible update orders of all blocks occur, we
	show that they suffice to cover all possible feasible
	solutions. We provide a way to construct a valid update order accordingly.
	With these intuitions in mind, we now present a rigorous 
	analysis. Let $\pi_{S_1}=(a_{1},\ldots,a_{\ell_1})$ and
	$\pi_{S_2}=(a'_1,\ldots,a'_{\ell_2})$ be permutations of sets $S_1$ and
	$S_2$. We define the \emph{core}
	of $\pi_{S_1}$ and $\pi_{S_2}$ as $core(\pi_{S_1},\pi_{S_2}) :=
	S_1\cap S_2$. We say that two permutations $\pi_1$ and $\pi_2$ 
	are \emph{consistent}, $\pi_1
	\consistent \pi_2$, if there is a permutation $\pi$ of symbols of
	$core(\pi_1,\pi_2)$ such that $\pi$ is a subsequence of both $\pi_1$
	and $\pi_2$.
	
	The \textbf{Dependency Graph} is 
	a labelled graph defined recursively as follows. The
	dependency graph of a single permutation $\pi=(a_1,\ldots,a_\ell)$, denoted by $G_{\pi}$, is a 
	directed path
	$v_1,\ldots,v_{\ell}$, and the label of the vertex 
	$v_i\in V(G_{\pi})$ is the element
	$a$ with $\pi(a)=i$. We denote by $\Labels{G_{\pi}}$ the set of all 
	labels of $G_{\pi}$. 
	
	Let $G_{\Pi}$ be a dependency graph of the set of 
	permutations $\Pi$ and $G_{\Pi'}$ the dependency graph of
	the set $\Pi'$. Then, their union (by identifying the same vertices)
	forms the dependency graph $G_{\Pi\cup\Pi'}$ of the set
	$\Pi\cup\Pi'$. Note that such a dependency graph is not necessarily acyclic.
	
	We call a permutation $\pi$ of blocks of a subset $\blockset'\subseteq\blockset$ 
	\emph{congestion free}, if the following holds: 
	it is possible to update the blocks
	in $\pi$ in the graph $G_{\blockset}$ (the graph on the union of blocks in $\blockset$), in
	order of their appearance in $\pi$, without violating any edge
	capacities in $G_{\blockset}$. Note that we do not respect all conditions
	of our \emph{Consistency Rule} (definition~\ref{def:consistencyrule}) here.

	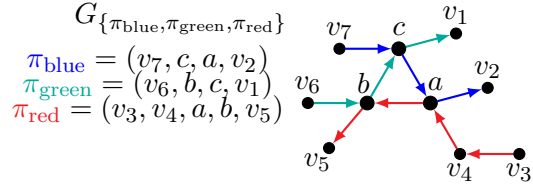
\begin{figure}[h!]
		\begin{center}
			\begin{tikzpicture}[scale=0.6]
			\tikzset{>=latex} 
			
			\node (c1) [] {};
			\node (c2) [position=180:55mm from c1] {};
			
			\node (s1) [position=90:5.5mm from c2] 
			{$\textcolor{blue}{\pi_{\operatorname{blue}}}=(v_7,c,a,v_2)$};		
			\node (s2) [position=0:0mm from c2] 
			{$\textcolor{JungleGreen}{\pi_{\operatorname{green}}}=(v_6,b,c,v_1)$};
			\node (s3) [position=270:5.5mm from c2] 
			{$\textcolor{Red}{\pi_{\operatorname{red}}}=(v_3,v_4,a,b,v_5)$};
			
			\node (graph) [position=163:50mm from c1] {$G_{\left\{ 
			\pi_{\operatorname{blue}},\pi_{\operatorname{green}},\pi_{\operatorname{red}} 
			\right\}}$};

			\node (a) [draw,circle,fill,position=330:8mm from c1,scale=0.5] {};
			\node (la) [position=75:4.5mm from a] {$a$};
			\node (b) [draw,circle,fill,position=210:8mm from c1,scale=0.5] {};
			\node (lb) [position=100:4.5mm from b] {$b$};
			\node (c) [draw,circle,fill,position=90:8mm from c1,scale=0.5] {};
			\node (lc) [position=90:4.5mm from c] {$c$};
			
			\node (v1) [draw,circle,fill,position=15:13mm from c,scale=0.4] {};
			\node (lv1) [position=90:4mm from v1] {$v_1$};
			\node (v2) [draw,circle,fill,position=15:13mm from a,scale=0.4] {};
			\node (lv2) [position=90:4mm from v2] {$v_2$};
			\node (v4) [draw,circle,fill,position=300:13mm from a,scale=0.4] {};
			\node (lv4) [position=270:4mm from v4] {$v_4$};
			\node (v3) [draw,circle,fill,position=0:13mm from v4,scale=0.4] {};
			\node (lv3) [position=270:4mm from v3] {$v_3$};
			\node (v5) [draw,circle,fill,position=230:13mm from b,scale=0.4] {};
			\node (lv5) [position=230:4mm from v5] {$v_5$};
			\node (v6) [draw,circle,fill,position=180:13mm from b,scale=0.4] {};
			\node (lv6) [position=90:4mm from v6] {$v_6$};
			\node (v7) [draw,circle,fill,position=180:13mm from c,scale=0.4] {};
			\node (lv7) [position=90:4mm from v7] {$v_7$};

			\begin{pgfonlayer}{bg}   
			
			\draw [thick,->,color=blue] (v7) to (c);
			\draw [thick,->,color=blue] (c) to (a);
			\draw [thick,->,color=blue] (a) to (v2);
			
			\draw [thick,->,color=JungleGreen] (v6) to (b);
			\draw [thick,->,color=JungleGreen] (b) to (c);
			\draw [thick,->,color=JungleGreen] (c) to (v1);
			
			\draw [thick,->,color=Red] (v3) to (v4);
			\draw [thick,->,color=Red] (v4) to (a);
			\draw [thick,->,color=Red] (a) to (b);
			\draw [thick,->,color=Red] (b) to (v5);
			
			\end{pgfonlayer}

			\end{tikzpicture}	
		\end{center}
		\vspace{-1em}
		\caption{\emph{Example:} The dependency graph of three
			pairwise consistent permutations
			$\textcolor{blue}{\pi_{\operatorname{blue}}}$,
			$\textcolor{JungleGreen}{\pi_{\operatorname{green}}}$ and
			$\textcolor{Red}{\pi_{\operatorname{red}}}$. Each pair of
			those permutation has exactly one vertex in common and with
			this the cycle $(a,b,c)$ is created. With such cycles being
			possible a dependency graph does not necessarily
			contain sink vertices. To get rid of them, we certainly need some more refinements.}
		\label{fig:exdependency1}
	\end{figure}
	
	In the approach we are taking, one of the main advantages we have is the nice properties of 
	blocks when 
	it comes to updating.
	The following algorithm formalizes the procedure already described in 
	\Cref{lem:updatewholeblock}.
	The correctness follows directly from said lemma.
	Let $P=(F^o,F^u)$ be a given flow pair.
	
	\begin{algorithm}\textbf{Update a Free Block $b$}\label[algorithm]{alg:updatefreeblock}
		\begin{enumerate}
			\item Resolve $(v,P)$ for all $v\in F^u \cap b-\Start{b}$.
			\item Resolve $(\Start{b},P)$.
			\item Resolve $(v,P)$ for all $v\in (b-F^u)$.
			\item For any edge in $E(b\cap F^u)$ check whether $d_{F^u}$ together with the other 
			loads 
			on $e$ exceed $c(e)$. If so output: \emph{Fail}.
		\end{enumerate}
	\end{algorithm}
	
	\begin{lemma}\label[lemma]{lem:computingcongestionfree}
		Let $\pi$ be a permutation of the set $\blockset_1\subseteq
		\blockset$. Whether $\pi$ is congestion free can be determined
		in time $O(k\cdot\sizeof{G})$.
	\end{lemma}
	\begin{proof}
		In the order of $\pi$, perform~\Cref{alg:updatefreeblock}. If
		it fails, i.e., if it violates congestion freedom for some edges, $\pi$
		is not a congestion free permutation. The running time
		of~\Cref{alg:updatefreeblock} is in $O(\sizeof{b})$ for a
		block $b$, hence the overall running time is bounded above by:
		\[
		\sum_{b\in \blockset_1} \sizeof{b}= \sum_{i=1}^{k}\sum_{\substack{b\in \blockset_1\\ b\text{ is
					an }i\text{-block}}}\sizeof{b} \le k\cdot \sizeof{G}.
		\]
		
	\end{proof}
	
	The smaller relation defines a total order on all blocks in $G$. Let
	$\blockset=\{b_1,\ldots,b_{\sizeof{\blockset}}\}$ and suppose the order is
	$b_1<\ldots<b_{\sizeof{\blockset}}$.
	\smallskip
	
	We define an auxiliary graph $H$ which will help us find a suitable
	dependency graph for our network. We first provide some high-level definitions 
	relevant to the construction of the graph $H$ only. 
	Exact definitions will follow in the construction of $H$, and will be used
	throughout the rest of this section.
	
	Recall that $\blockset$ is the set of all blocks in $G$. We define
	another set of blocks $\blockset'$ which is initialized as $\blockset$;
	the construction of $H$ is iterative, and in each iteration, we eliminate
	a block from $\blockset'$. At the end of the construction of $H$, $\blockset'$ is empty.
	For every block $b\in \blockset'$, we also define the set $\TBs{b}$ of
	blocks which touch the block $b$. Another set which is defined
	for every block $b$ is the set $\BTSset{b}$; this set actually
	corresponds to a set of vertices, each of which corresponds to a
	valid congestion free permutation of blocks in $\TBs{b}$. Clearly if
	$\TBs{b}$ does not contain any congestion-free permutation, then $\BTSset{b}$ is an
	empty set. As we already mentioned, every vertex $v\in \BTSset{b}$
	comes with a \textbf{label} which corresponds to some congestion-free
	permutation of elements of $\TBs{b}$. We denote that permutation with $\Label{(v)}$.
	
	\smallskip
	
	\textbf{Construction of $H$: } We recursively construct a
	labelled graph $H$ from the blocks of $G$ as follows. 
	
	\begin{enumerate}[i]
		\item Set $H\coloneqq \emptyset$, $\blockset':=\blockset$, $\BTSsets:=\emptyset$.
		\item For $i\coloneqq 1,\ldots,\sizeof{\blockset}$ do
		\begin{enumerate}[1]
			\item []\label{def:tislessthank}
			\item Let $b\coloneqq b_{\sizeof{\blockset}-i+1}$.
			\item Let $\TBs{b}\coloneqq \{b'_1,\ldots,b'_t\}$ be the set of blocks in $\blockset'$ touched 
			by $b$.
			\item Let $\pi\coloneqq \{\pi_1,\ldots,\pi_{\ell}\}$ be the set of congestion
			free permutations of $\TBs{b}$.
			\item Set $\BTSset{b}\coloneqq \emptyset$.
			\item For $i\in[\ell]$ create a vertex $v_{\pi_i}$ with
			$\Label(v_{\pi_i})=\pi_i$ and set $\BTSset{b}\coloneqq \BTSset{b}\cup v_{\pi_i}$.
			\item Set $H\coloneqq H\cup \BTSset{b}$.
			\item Add edges between all pairs of vertices in $H[\BTSset{b}]$.
			\item Add an edge between every pair of vertices $v\in H[\BTSset{b}]$ and $u\in
			V(H) - \BTSset{b}$ if the labels of $v$ and $u$ are
			inconsistent.
			\item \label{def:removeb} Set $\blockset':=\blockset' - b$.
		\end{enumerate}
	\end{enumerate}
	
	\begin{figure}[h!]
		\begin{center}
			\begin{tikzpicture}[scale=0.8]
			\tikzset{>=latex} 
			
			\node (c1) [] {};
			
			\node (before) [position=180:25mm from c1,inner sep=0pt] {$\dots$};
			\node (V1center) [position=180:14mm from c1,inner sep=0pt] {};
			\node (V2center) [position=0:0mm from c1,inner sep=0pt] {};
			\node (between) [position=0:17mm from c1,inner sep=0pt] {$\dots$};
			\node (V3center) [position=0:34mm from c1,inner sep=0pt] {};
			\node (after) [position=0:48mm from c1,inner sep=0pt] {$\dots$};
			
			\node (V1label) [position=90:26mm from V1center,inner sep=0pt] {$\BTSset{b_i}$};
			\node (V2label) [position=90:21mm from V2center,inner sep=0pt] {$\BTSset{b_{i+1}}$};
			\node (V3label) [position=90:26mm from V3center,inner sep=0pt] {$\BTSset{b_j}$};
			
			\node (V1) [position=0:0 from V1center,draw,line width=1pt,minimum height=70pt,minimum 
			width=21pt] {};
			\node (V2) [position=0:0 from V2center,draw,line width=1pt,minimum height=70pt,minimum 
			width=21pt] {};
			\node (V3) [position=0:0 from V3center,draw,line width=1pt,minimum 
			height=70pt,minimum width=21pt] {};
			
			\node (v11) [position=270:12mm from V1center,inner sep=0pt] {};
			\node (v12) [position=270:8mm from V1center,inner sep=0pt] {};
			\node (v13) [position=270:4mm from V1center,inner sep=0pt] {};
			\node (v14) [position=270:0mm from V1center,inner sep=0pt] {};
			\node (v15) [position=90:4mm from V1center,inner sep=0pt] {};
			\node (v16) [position=90:8mm from V1center,inner sep=0pt] {};
			\node (v17) [position=90:12mm from V1center,inner sep=0pt] {};
			
			\node (v21) [position=270:12mm from V2center,inner sep=0pt] {};
			\node (v22) [position=270:8mm from V2center,inner sep=0pt] {};
			\node (v23) [position=270:4mm from V2center,inner sep=0pt] {};
			\node (v24) [position=270:0mm from V2center,inner sep=0pt] {};
			\node (v25) [position=90:4mm from V2center,inner sep=0pt] {};
			\node (v26) [position=90:8mm from V2center,inner sep=0pt] {};
			\node (v27) [position=90:12mm from V2center,inner sep=0pt] {};
			
			\draw (v17) to (v27);
			\draw (v17) to (v26);
			\draw (v17) to (v24);
			\draw (v14) to (v25);
			\draw (v15) to (v22);
			\draw (v11) to (v23);
			
			\node (v31) [position=270:12mm from V3center,inner sep=0pt] {};
			\node (v32) [position=270:8mm from V3center,inner sep=0pt] {};
			\node (v33) [position=270:4mm from V3center,inner sep=0pt] {};
			\node (v34) [position=270:0mm from V3center,inner sep=0pt] {};
			\node (v35) [position=90:4mm from V3center,inner sep=0pt] {};
			\node (v36) [position=90:8mm from V3center,inner sep=0pt] {};
			\node (v37) [position=90:12mm from V3center,inner sep=0pt] {};
			
			\draw [color=red,line width=0.6pt,bend left=26] (v17) to (v37);
			\draw [color=red,line width=0.6pt,bend right=35] (v11) to (v32);
			
			\draw [color=red,line width=0.6pt,bend left=20] (v26) to (v37);
			\draw [color=red,line width=0.6pt,bend left=20] (v24) to (v35);
			\draw [color=red,line width=0.6pt,bend left=20] (v22) to (v31);
			
			\node (gs2) [position=292:28mm from c1,inner sep=0pt] {};
			\node (gs1) [position=180:36mm from gs2,inner sep=0pt] {};
			\node (gs3) [position=0:36mm from gs2,inner sep=0pt] {};	
			
			\node (s2) [position=270:5mm from gs2,draw,minimum width=70pt,minimum 
			height=10pt,color=blue,line width=0.7pt] {};
			\node (s1) [position=270:5mm from gs1,draw,minimum width=70pt,minimum 
			height=10pt,color=blue,line width=0.7pt] {};
			\node (s3) [position=270:5mm from gs3,draw,minimum width=70pt,minimum 
			height=10pt,color=blue,line width=0.7pt] {};			
			
			\begin{pgfonlayer}{bg}   
			
			\node (b1) [draw,circle,line width=0.9,color=blue,minimum width=6pt,inner 
			sep=3.7pt,position=0:0mm from v17] {};
			\node (b2) [draw,circle,line width=0.9,color=blue,minimum width=6pt,inner 
			sep=3.7pt,position=0:0mm from v26] {};
			\node (b3) [draw,circle,line width=0.9,color=blue,minimum width=6pt,inner 
			sep=3.7pt,position=0:0mm from v37] {};
			
			\end{pgfonlayer}
			
			\draw [->,line width=0.7pt,color=blue,bend right=20] (b1) to (gs1);
			\draw [->,line width=0.7pt,color=blue,bend left=20] (b2) to (gs2);
			\draw [->,line width=0.7pt,color=blue,bend left=20] (b3) to (gs3);
			
			\node (s1d1) [position=0:0mm from s1] {$\dots$};
			\node (s2d1) [position=0:0mm from s2] {$\dots$};
			\node (s3b1) [position=0:0mm from s3,draw,ellipse,line width=0.5pt,color=red] 
			{$\textcolor{black}{c\dots d}$};
			\node (s2b1) [position=180:8.8mm from s2d1,draw,ellipse,line width=0.5pt,color=red] 
			{$\textcolor{black}{d\dots c}$};
			\node (s1b1) [position=180:8.8mm from s1d1,draw,ellipse,line width=0.5pt,color=red] 
			{$\textcolor{black}{d\dots c}$};
			\node (s2b2) [position=0:8.6mm from s2d1,draw,ellipse,line width=0.5pt] 
			{$\textcolor{black}{a\dots b}$};
			\node (s1b2) [position=0:8.6mm from s1d1,draw,ellipse,line width=0.5pt] 
			{$\textcolor{black}{b\dots a}$};

			\draw [line width=0.8pt,color=red,bend right] (s1b1) to (s3b1);
			\draw [line width=0.8pt,color=red,bend right] (s2b1) to (s3b1);
			
			\draw [line width=0.8pt,bend right=60] (s1b2) to (s2b2);

			\end{tikzpicture}	
		\end{center}
		\vspace{-1em}
		\caption{\emph{Example:} The graph $H$ consists of vertex sets
			$\BTSset{b_i}$, $i\in[\sizeof{\blockset}]$, where each such partition contains all
			congestion free sequences of the at most $k$ iteratively chosen
			touching blocks. In the whole graph, we then create edges between the
			vertices of two such partitions if and only if the corresponding
			sequences are inconsistent with each other, as seen in the three
			highlighted sequences.
			Later we will distinguish between such edges connecting
			vertices of neighbouring partitions (w.r.t.~the topological
			order of their corresponding blocks), $\BTSset{b_i}$ and
			$\BTSset{b_{i+1}}$, and partitions that are further away, $\BTSset{b_i}$
			and $\BTSset{b_j}$. Edges of the latter type, depicted as red in the
			figure, are called long edges and will be deleted in the
			trimming process of $H$.}
		\label{fig:H1}
	\end{figure}
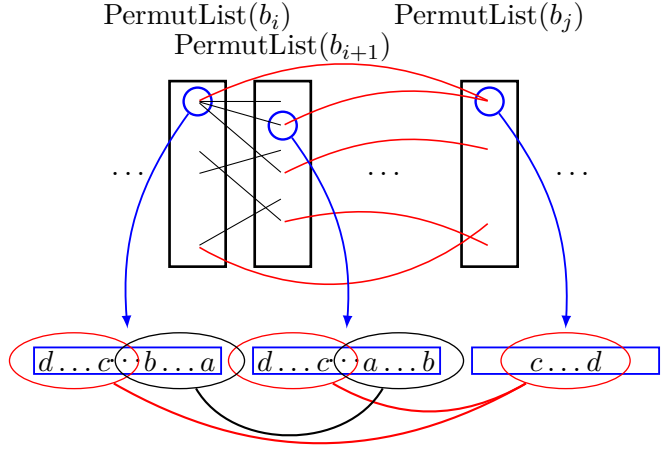
	
	\begin{lemma}\label[lemma]{lem:limitsizeofpermutations}
		For Item~(\ref{def:tislessthank}) of the construction of $H$, $t\le k$ holds.
	\end{lemma}
	\begin{proof}
		Suppose for the sake of contradiction that $t$ is bigger than $k$. 
		So there are $j$-blocks $b,b'$ (where $b_{\sizeof{\blockset}-i+1}$ corresponds to a flow pair 
		different from $j$) that touch $b_{\sizeof{\blockset}-i+1}$. But then one of
		$\Start{b}$ or $\Start{b'}$ is strictly larger than $\Start{b_{\sizeof{\blockset}-i+1}}$. This
		contradicts our choice of $b_{\sizeof{\blockset}-i+1}$ in that we deleted
		larger blocks from $\blockset'$ in Item~(\ref{def:removeb}).
	\end{proof}
	
	\begin{lemma}[Touching Lemma]\label{lem:touchesomethinginmiddle}
		Let $b_{j_1},b_{j_2},b_{j_3}$ be three blocks (w.r.t.~$<$)
		where $j_1<j_2<j_3$. Let $b_z$ be another block such that $z \notin \left\{ j_1,j_2,j_3 \right\}$.
		If in the process of constructing $H$, $b_z$ is in the touch list of both
		$b_{j_1}$ and $b_{j_3}$, then it is also in the touch list of $b_{j_2}$.
	\end{lemma}
	\begin{proof}
		Let us suppose that $\Start{b_{j_1}}\neq\Start{b_{j_2}}\neq \Start{b_{j_3}}$.
		We know that $\Start{b_z}\prec \Start{b_{j_1}}$ as otherwise, in the
		process of creating $H$, we eliminate $b_z$ before we process
		$b_{j_1}$: it would hence not appear in the touch list of
		$b_{j_1}$. As $b_z\touches b_{j_3}$,
		there is a vertex $v\in b_z$ where $\Start{b_{j_3}}\prec v$. But by
		our choice of elimination order: $\Start{b_{j_2}}\prec
		\Start{b_{j_3}}\prec v\prec\End{b_z}$, and on the other hand:
		$\Start{b_z}\prec \Start{b_{j_1}}\prec \Start{b_{j_2}}$. Thus, 
		$\Start{b_z}\prec \Start{b_{j_2}}\prec \End{b_z}$, and therefore
		$b_z$ touches $b_{j_2}$. 
		If some of the start vertices are the same, a similar case distinction 
		applies.
	\end{proof}
	
	\begin{figure}[h!]
		\begin{center}
			\begin{tikzpicture}[scale=0.6]
			\tikzset{>=latex} 
			
			\node (c1) [] {};
			
			\node (e1center) [position=180:20mm from c1,inner sep=0pt] {};
			\node (e2center) [position=0:0mm from c1,inner sep=0pt] {};
			\node (e3center) [position=0:20mm from c1,inner sep=0pt] {};
			
			\node (e1label) [position=150:16mm from e1center,inner sep=0pt] 
			{$\pi_{b_{\operatorname{blue}}}$};
			\node (e2label) [position=90:24mm from e2center,inner sep=0pt] 
			{$\pi_{b_{\operatorname{green}}}$};
			\node (e3label) [position=30:16mm from e3center,inner sep=0pt] 
			{$\pi_{b_{\operatorname{red}}}$};
			
			\node (e1) [position=0:0 from e1center,draw,line width=1pt,ellipse,minimum 
			height=70pt,minimum width=21pt,color=blue] {};
			\node (e2) [position=0:0 from e2center,draw,line width=1pt,ellipse,minimum 
			height=70pt,minimum width=21pt,color=JungleGreen] {};
			\node (e3) [position=0:0 from e3center,draw,line width=1pt,ellipse,minimum 
			height=70pt,minimum width=21pt,color=red] {};
			
			\node (t1) [position=240:26mm from e1center] {};
			\node (t2) [position=300:26mm from e3center] {};
			\node (topo) [position=270:28mm from e2center] {topological order};
			
			\node(bg1) [position=90:7mm from e1center,inner sep=0pt] {};
			\node(gg1) [position=90:10mm from e2center,inner sep=0pt] {};
			\node(gg2) [position=270:10mm from e2center,inner sep=0pt] {};
			\node(rg1) [position=0:0mm from e3center,inner sep=0pt] {};
			
			\node(ra) [position=90:3mm from rg1,inner sep=0pt] {$\textcolor{red}{a}$};
			\node(rb) [position=270:3mm from rg1] {$\textcolor{red}{b}$};
			\node(rd1) [position=90:15mm from rg1] {$\textcolor{red}{\vdots}$}; 
			\node(rd2) [position=270:12mm from rg1] {$\textcolor{red}{\vdots}$}; 
			\node(erg1) [position=0:0mm from rg1,draw,dashed,ellipse,minimum height=30pt,minimum 
			width=8pt] {};
			
			\node(bc) [position=90:3mm from bg1] {$\textcolor{blue}{c}$};
			\node(ba) [position=270:3mm from bg1,inner sep=0pt] {$\textcolor{blue}{a}$};
			\node(bd1) [position=270:13mm from bg1] {$\textcolor{blue}{\vdots}$}; 
			\node(ebg1) [position=0:0mm from bg1,draw,dashed,ellipse,minimum height=30pt,minimum 
			width=8pt] {};
			
			\node(ga1) [position=90:3mm from gg1,inner sep=0pt] {${a}$};
			\node(gb) [position=270:3mm from gg1] {$\textcolor{JungleGreen}{b}$};
			\node(egg1) [position=0:0mm from gg1,draw,dashed,ellipse,minimum height=30pt,minimum 
			width=8pt] {};
			
			\node(gc) [position=90:3mm from gg2] {$\textcolor{JungleGreen}{c}$};
			\node(ga2) [position=270:3mm from gg2,inner sep=0pt] {${a}$};
			\node(egg2) [position=0:0mm from gg2,draw,dashed,ellipse,minimum 
			height=30pt,minimum width=8pt] {};
			
			\draw [->,bend right=20] (ba) to (ga2);
			\draw [->,bend right=20] (ra) to (ga1);
			
			\begin{pgfonlayer}{bg}  
			
			\draw [->] (t1) to (t2);

			\end{pgfonlayer}

			\end{tikzpicture}	
		\end{center}
		\vspace{-1em}
		\caption{\emph{Example:} Select one of the permutations of
			length at most $k$ from every $\BTSset{b}$. These
			permutations obey the~\nameref{lem:touchesomethinginmiddle}. Taking the three 
			permutations from the example in Figure~\ref{fig:exdependency1}, we can see that 
			the~\nameref{lem:touchesomethinginmiddle} forces $a$ to be in the green permutation as 
			well. Assuming consistency, this would mean $a$ to come \emph{before} $b$ and 
			\emph{after} $c$. Hence $a<_{\pi_{\operatorname{green}}}b$ and 
			$b<_{\pi_{\operatorname{green}}}a$, a contradiction. So if our permutations are derived 
			from 
			$H$ and are consistent, we will show that cycles cannot occur in their dependency graph.}
		\label{fig:extouchingdependency1}
	\end{figure}

	For an illustration of the property described in the~\nameref{lem:touchesomethinginmiddle}, see 
	Figure~\ref{fig:extouchingdependency1}: it refers to the dependency
	graph of Figure \ref{fig:exdependency1}. This example also points out
	the problem with directed cycles in the dependency graph and the
	property of the~\nameref{lem:touchesomethinginmiddle}.
	
	We prove some lemmas in regard to the dependency graph of elements of $H$,
	to establish the base of the inductive proof for~\Cref{lem:dependencygraphacyclic}.
	
	We begin with a simple observation on the fact that a permutation $\pi$ induces a total order on 
	the elements of $S$.
	
	\begin{observation}\label{obs:onepermutation}
		Let $\pi$ be a permutation of a set $S$. Then the dependency graph
		$G_\pi$ does not contain a cycle.
	\end{observation}
	
	\begin{lemma}\label[lemma]{lem:twopermutation}
		Let $\pi_1,\pi_2$ be permutations of sets $S_1,S_2$ such that
		$\pi_1,\pi_2$ are consistent. Then the dependency graph $G_{\pi_1\cup\pi_2}$ is acyclic.
	\end{lemma}
	\begin{proof}
		For the sake of contradiction suppose there is a cycle $C$ in
		$G_{\pi_1\cup\pi_2}$. By~\Cref{obs:onepermutation} this cycle must contain vertices 
		corresponding to elements of both $S_1$ and $S_2$. Let $a$ be the least element of $S_1$ 
		with
		respect to $\pi_1$ such that $v_a\in V(C)$. As $C$ is a cycle there is a vertex $v_b$ with 
		$b\in S_1\cup S_2$ such that the edge $(v_b,v_a)$ is an edge of $C$. By our choice of $a$, 
		$b$ is not contained in $S_1$. Hence, since the edge $(v_b,v_a)$ exists, $a\in S_1\cap S_2$. 
		Similarly we can consider the least element $c\in S_2$ in $C$ and its predecessor $d\in 
		S_1\setminus S_2$ along the cycle. Again the edge $(v_d,v_c)$ exists and thus $c\in S_1\cap 
		S_2$. Now we have $d<a$ in $\pi_2$, but $a<d$ in $\pi_1$ contradicting the consistency of 
		$\pi_1$ and $\pi_2$.
	\end{proof}
	
	In the next lemma, we need a closure of the dependency graph of permutations which we
	define as follows.
	\begin{definition}[Permutation Graph Closure]\label{def:closurepermutation}
		The \emph{Permutation Graph Closure}, or simply \emph{closure}, of a permutation $\pi$ is 
		the graph $G^+_{\pi}$
		obtained from taking the transitive closure of $G_{\pi}$, i.e.~its
		vertices and labels are the same as $G_{\pi}$ and there is an edge $(u,v)$ in
		$G^+_{\pi}$ if there is a path starting at $u$ and ending at $v$ in
		$G_{\pi}$. Similarly the \emph{Permutation Graph Closure} of a set of permutations
		$\Pi=\{\pi_1,\ldots,\pi_n\}$ is the graph obtained by taking the union of
		$G^+_{\pi_i}$'s (for $i\in [n]$) by identifying vertices of the same label.
	\end{definition}
	
	In the above definition note that if $\Pi$ is a set of permutations
	then $G_\Pi\subseteq G^+_\Pi$.
	
	The following lemma
	generalizes~\Cref{lem:twopermutation,obs:onepermutation}
	and uses them as the base of its inductive proof. 
	
	\begin{lemma}\label[lemma]{lem:dependencygraphacyclic}
		Let $I=\{v_{\pi_1},\ldots,v_{\pi_\ell}\}$ be an independent set in $H$. Then the
		dependency graph $G_\Pi$, for $\Pi=\{\pi_1,\ldots,\pi_\ell\}$, is
		acyclic.
	\end{lemma}
	\begin{proof}
		Instead of working on $G_{\Pi}$, we can work on its closure
		$G^+_{\Pi}$ as defined above. First we
		observe that every edge in $G_{\Pi}$ also appears in $G^+_{\Pi}$, so
		if there is a cycle in $G_{\Pi}$, the same cycle exists in $G^+_{\Pi}$.
		
		We prove that there is no cycle in
		$G^+_\Pi$. By~\Cref{lem:twopermutation} and~\Cref{obs:onepermutation} there is no
		cycle of length at most~$2$ in $G^+_\Pi$; otherwise there is a
		cycle in $G_{\Pi}$ which consumes at most two consistent permutations.
		
		For the sake of contradiction, suppose $G^+_{\Pi}$ has a cycle and let
		$C=(a_1,\ldots,a_n)\subseteq G^+_\Pi$ be the shortest cycle in
		$G^+_{\Pi}$. By~\Cref{lem:twopermutation} and~\Cref{obs:onepermutation} we
		know that $n\geq 3$.
		
		In the following, because we work on a cycle $C$, whenever we write any index
		$i$ we consider it w.r.t.~its cyclic order on $C$, in fact $i\mod
		\sizeof{C}+1$. So for example, $i=0$ and $i=n$ are identified as the same indices;
		similarly for $i=n+1,i=1$, etc.
		
		Recall the construction of the dependency graph where every vertex
		$v\in C$ corresponds to some block $b_v$. In
		the remainder of this proof we do not distinguish between the vertex $v$ and the
		block $b_v$.
		
		Let $\pi_v$ be the label of a given vertex $v\in I$. 
		For each edge $e=(a_i,a_{i+1}) \in C$, there is a permutation
		$\pi_{v_i}$ such that $(a_i,a_{i+1})$ is a subsequence of $\pi_{v_i}$
		and additionally the vertex $v_i$ is in the set $I$. So there is a
		block $b^i$ such that $\pi_{v_i}$ is a permutation of the set $\TBs{b^i}$.
		
		The edge $e=(a_i,a_{i+1})$ is said to \emph{represent} $b^i$, and we call it the representative 
		of $\pi_{v_i}$. For each
		$i$ we fix one block $b^i$ which is represented by the edge $(a_i,a_{i+1})$ (note that one 
		edge can represent many
		blocks, but here we fix one of them). We define the set of those blocks as
		$B^I=\{b^1,\ldots,b^\ell\}$ and state the following claim.
		
		\begin{Claim}\label{clm:btslength}
			For every two distinct vertices $a_i,a_j\in C$, either there is no block $b\in B^I$
			such that $a_i,a_j\in \TBs{b}$ or if $a_i,a_j\in \TBs{b}$ then
			$(a_i,a_j)$ or $(a_j,a_i)$ is an edge in $C$. Additionally $\sizeof{B^I}=\sizeof{C}$.
		\end{Claim}
		\begin{ClaimProof}
			Suppose there is a block $b\in B^I$ such that
			$a_i,a_j\in \TBs{b}$. Then in $E(G^+_{\Pi})$ there is an edge
			$e_1=(a_i,a_j)$ or $e_2=(a_j,a_i)$. If either of
			$e_1,e_2$ is an edge in $C$ then we are done. Otherwise if $e_1\in
			E(G^+_{\Pi})$ then the cycle on the vertices $a_1,\ldots,a_i,a_j,\ldots,a_n$ is
			shorter than $C$ and if $e_2\in E(G^+_{\Pi})$ then the cycle on
			the vertices $a_i,\ldots,a_j$ is shorter than $C$. Both cases
			contradict the assumption that $C$ is the shortest cycle in
			$G^+_{\Pi}$.
			For the second part of the claim it is clear that $\sizeof{B^I}\le \sizeof{C}$, 
			on the other hand if both endpoints of an edge $e=(a_i,a_{i+1})\in C$
			appear in $\TBs{b}$ and $\TBs{b'}$ for two different blocks $b,b'\in B^I$
			then, by our choice of the elements of $B^I$, at least one of them (say $b$) has a 
			representative $e'\neq e$. But, then there is a vertex
			$a_j\in V(e')$ such that $a_j\neq a_i, a_j\neq a_{i+1}$. But by the
			first part this
			cannot happen, so we have $\sizeof{C}\le
			\sizeof{B^I}$ and the second part of the claim follows.
		\end{ClaimProof}
		
		By the above claim we have $\ell=n$.
		W.l.o.g. suppose $b^1< b^2<\ldots<b^n$. There is an $i\in [n]$ such
		that $(a_{i-1},a_i)$ represents $b^1$, we fix this $i$.
		
		\begin{Claim}\label{clm:previsb2}
			If $(a_{i-1},a_i)$ represents $b^1$ then $(a_{i-2},a_{i-1})$ represents $b^2$.
		\end{Claim}
		\begin{ClaimProof}
			By Claim~\ref{clm:btslength} there
			is a block $b^t$ represented by $(a_{i-2},a_{i-1})$. We also have $b^1 < b^2 \le b^t$ hence 
			by the~\nameref{lem:touchesomethinginmiddle}, $a_{i-1}$ appears in $\TBs{b^2}$.
			But then by Claim~\ref{clm:btslength} either $a_{i+1}$ is in
			$\TBs{b^2}$ or $a_{i-2}\in \TBs{b^2}$, by the former case we have
			$b^1=b^2$ which is a contradiction to the assumption that
			$b^1<b^2$. In the latter case we have $t=2$ which proves
			the claim.
		\end{ClaimProof}
		
		Similarly we can prove the endpoints of the edges, that have $a_i$ as their head, are in $b^2$.
		
		\begin{Claim}\label{clm:nextisb2}
			If $(a_{i-1},a_i)$ represents $b^1$ then $(a_{i},a_{i+1})$ represents $b^2$.
		\end{Claim}
		\begin{ClaimProof}
			By Claim~\ref{clm:btslength} there
			is a block $b^t$ such that $(a_{i},a_{i+1})$ represents $b^t$. We also have $b^1 < b^2 \le 
			b^t$ thus by
			the~\nameref{lem:touchesomethinginmiddle}, $a_{i}$ appears in $\TBs{b^2}$.
			But, then by Claim~\ref{clm:btslength} either $a_{i-1}$ or $a_{i+1}$ is in
			$\TBs{b^2}$. In the former case we have
			$b^1=b^2$ which is a contradiction to the assumption that
			$b^1<b^2$. In the latter case we have $t=2$ which proves
			the claim.
		\end{ClaimProof}
		By Claims~\ref{clm:previsb2}~and~\ref{clm:nextisb2} we have that both
		$(a_{i-2},a_{i-1})$ and $(a_{i},a_{i+1})$ represent $b^2$
		hence by Claim~\ref{clm:btslength} they are the same edge. Thus
		there is a cycle on the vertices $a_{i-1},a_i$ in $G^+_{\Pi}$ and this
		gives a cycle in $G_{\Pi}$ on at most~$2$ consistent permutations which is a
		contradiction according to~\Cref{lem:twopermutation}.
	\end{proof} 
	
	The following lemma establishes the link between independent sets in 
	$H$ and feasible update sequences of the corresponding update flow network $G$.
	
	\begin{lemma}\label[lemma]{lem:blocksandsets}
		There is a feasible sequence of updates for an update network $G$ on $k$
		flow pairs, if and only if there is an independent set of size
		$\sizeof{\blockset}$ in $H$. Additionally if the independent set $I\subseteq V(H)$ of size
		$\sizeof{\blockset}$ together with its vertex labels are given, then there is an algorithm which 
		can compute
		a feasible sequence of updates for $G$ in $O(k\cdot \sizeof{G})$.
	\end{lemma}
	\begin{proof}
		First we prove that if there is a sequence of feasible updates $\updatesequence$,
		then there is an independent set of size $\sizeof{\blockset}$ in $H$. Suppose 
		$\updatesequence$
		is a feasible sequence of updates of blocks. For a block $b$,
		recall that $\TBs{b}=\{b'_1,\ldots,b'_\ell\}$ is the set of
		remaining (not yet processed) blocks that touch
		$b$. Let $\pi_b$ be the reverse order of updates of blocks in $\TBs{b}$
		w.r.t.~$\updatesequence$. In fact, if $\updatesequence$ updates $b'_1$ first, then $b'_2$, 
		then
		$b_3',\ldots,b'_\ell$, then $\pi_b=b'_\ell \ldots b'_1$. 
		
		For every two blocks $b,b'\in I$, we have $\pi_b\consistent
		\pi_{b'}$.
		From every set of vertices $\BTSset{b}$, for $b\in B$, let
		$v^b_i$ be a vertex such that $\Label(v^b_i)$ is a subsequence of
		$\pi_b$. Recall that, the labels of vertices in $\BTSset{b}$ are all
		possible congestion free permutations of blocks that touch $b$ in the remaining set of blocks 
		$\blockset'$ during the construction of $H$. So the vertex $v^b_i$ exists. Put $v^b_i$ in $I$. 
		The labels of every
		pair of vertices in $I$ are consistent, as their super-sequences were
		consistent, so $I$ is an independent set and furthermore $\sizeof{ I}=\sizeof{\blockset}$.

		For the other direction, suppose there is an 
		independent set of vertices $I$ of size $\sizeof{\blockset}$ in $H$. It is
		clear that for every block $b\in
		\blockset$, there is exactly one vertex $v_b\in
		I\cap\BTSset{b}$.
		
		Let us define the dependency graph of the set of labels (permutations) $\Pi=\left\{ 
		\Label{(v_b)} ~|~
		b\in \blockset,v_b\in I\right\}$ as the dependency graph \emph{$D:=G_{\Pi}$}.
		$I$ is an independent set and thus every pair of labels of
		vertices in $I$ are consistent, hence by
		~\Cref{lem:dependencygraphacyclic} we know that $D$ is a DAG,
		and thus it has a sink vertex.
		
		We update blocks which correspond to
		sink vertices of $D$ in parallel by applying 
		\Cref{alg:updatefreeblock} and we remove those vertices from $D$
		after they are updated. Then we proceed
		recursively, until there is no vertex in $D$.
		We claim that this gives a feasible
		sequence of updates for all blocks. 
		
		Suppose there is a sink vertex whose corresponding block $b$ cannot be
		updated. There are two reasons preventing us from updating a block by ignoring the 
		Consistency 
		Rule: 
		\begin{enumerate}
			
			\item Its update stops the flow between some source and
			terminal. So afterwards there is no transient flow on the active edges.
			
			\item There is an edge $e\in E(b)$ which cannot
			be activated because this would imply routing along it and produce congestion.
			
		\end{enumerate}
		The first will never be the case by definition of \Cref{alg:updatefreeblock}. 
		So suppose there is such an edge $e$.
		Edge $e$ cannot be updated because some other
		blocks are incident to $e$ and currently route flows: updating 
		$b$ would violate a capacity
		constraint. There may be some blocks which are
		incident to $e$ but are not updated yet. These blocks would
		not effect the rest of our reasoning and we restrict ourselves
		to those blocks which have been updated already by our
		algorithm. Otherwise, if there is no such block, the label
		corresponding to $b$ is an invalid
		congestion free label. We will denote the set of the blocks preventing the update of $e$ by 
		$\blockset_e$.
		
		Suppose the blocks in $\blockset_e$ are updated in the order $b'_1,b'_2,\ldots,b'_\ell$ by 
		the above algorithm. Among $b,b'_1,\ldots,b'_{\ell}$, there is a block
		$b'$ which is the largest one (w.r.t.~$<$). In the construction of $H$, we know
		that $\BTSset{b'}\neq\emptyset$, as otherwise $I$ was not of size
		$\sizeof{\blockset}$. Suppose $v\in \BTSset{b'} \cap I$. In the iteration where we create
		$\BTSset{b'}$, $b'$ touches all blocks in
		$\{b,b'_1,\ldots,b'_\ell\}$, hence, in the $\Label(v)$, we
		have a subsequence $b''_1,\ldots,b''_{\ell+1}$ such that $b''_i\in
		\{b,b'_1,\ldots,b'_\ell\}$.
		
		We claim that the permutations $\pi_1=b''_1,\ldots,b''_{\ell+1}$ and
		$\pi_2=b'_1,\ldots,b'_\ell,b$ are exactly the same, which would contradict our
		assumption that $e$ cannot be updated: $\pi_1$ is a subsequence of
		the congestion free permutation $\Label(v)$.	
		Suppose $\pi_1\neq \pi_2$, then there are two 
		blocks $b'''_1,b'''_2$ with $\pi_1(b'''_1) <
		\pi_1(b'''_2)$ and $\pi_2(b'''_2)<\pi_2(b'''_1)$, then
		$\pi_1\not\consistent \pi_2$. Since both, 
		$b'''_2$ and $b'''_1$, will
		appear in $\Label(v)$, there is a directed path from
		$b'''_2$ to $b'''_1$ in $D$. Then our algorithm cannot choose $b'''_2$
		as a sink vertex before updating $b'''_1$: a
		contradiction. 
		
		At the end recall that we used 
		\Cref{alg:updatefreeblock} as a subroutine and this guarantees
		the existence of transient flow if we do not violate the
		congestion of edges, i.e. the algorithm does not return \emph{Fail} at any point.
		Hence, the sequence of updates we provided by deleting
		the sink vertices, is a valid sequence of updates if $I$ is an
		independent set of size $\sizeof{\blockset}$. 
		
		On the other hand, in the construction of $H$, all congestion free
		routings are already given and the runtime of \Cref{alg:updatefreeblock} is linear
		in the size of the dependency graph: If $I$ is given, the number of
		blocks is at most $k$ times larger than the original graph or
		$\sizeof{G_{\Pi}}=O(k\cdot \sizeof{G})$; therefore, we can compute
		the corresponding update sequence in $O(k\sizeof{G})$ as claimed.
	\end{proof}
	
	With~\Cref{lem:blocksandsets}, the update problem boils down to
	finding an independent set of size $\sizeof{\blockset}$ in $H$. However,
	this reduction does not suffice yet to solve our problem in polynomial
	time, as we will show next.
	
	Finding an independent set of size $\sizeof{\blockset}$ in $H$ is equivalent to finding
	a clique of size $\sizeof{\blockset}$ in its complement. The complement of $H$ is
	a $\sizeof{\blockset}$-partite graph where every partition has cardinality $\leq k!$. 
	In general, it is computationally hard to find such a clique. This is
	shown in the following lemma. Note that the lemma is not required
	for the analysis of our algorithm, but constitutes an independent result and serves to round off
	the discussion. 
	
	\begin{lemma}\label[lemma]{lem:cliqueinnpartitegraphishard}
		Finding an $m$-clique in an $m$-partite graph, where every partition has
		cardinality at most $3$, is NP-hard.
	\end{lemma}
	\begin{proof}
		We provide a polynomial time reduction from $3$-SAT. Let $C=C_1\wedge
		C_2\wedge\ldots \wedge C_m$ be an instance of $3$-SAT with~$n$ variables
		$X_1,\ldots,X_n$. We denote positive appearances of $X_i$ as a literal $x_i$ and negative 
		appearance as a literal
		$\bar{x}_i$ for $i\in [m]$. So we have at most $2n$ different literals
		$x_1,\ldots,x_n,\bar{x}_1,\ldots,\bar{x}_n$. Create an $m$-partite
		graph $G$ as follows. Set $G$ to be an empty graph.
		Let $C_i=\left\{ l_{i_1},l_{i_2},l_{i_3}\right\}$ be a clause for $i\in[m]$, then add vertices
		$v^i_{l_{i_1}},v^i_{l_{i_2}},v^i_{l_{i_3}}$ to $G$ as partition $p_i$. Note that $l_{i_1}=x_t$ or 
		$l_{i_1}=\bar{x}_t$ for
		some $t\in [n]$. Add an edge between each pair of
		vertices $v^i_x,u^j_y$ for $i,j\in [m], i\neq j$ if $x=x_t$ for some $t\in [n]$ and $y\neq
		\bar{x}_t$ or if $x=\bar{x}_t$ and $y\neq x_t$. It is clear that $G$ now is an $m$-partite 
		graph with exactly $3$ vertices in each partition.
		
		\begin{Claim} There is a satisfying assignment $\sigma$ for $C$ if, and only if,
			there is an $m$-clique in $G$.
		\end{Claim}
		\begin{ClaimProof}
			Define a vertex set $K=\emptyset$. Let $\sigma$ be a satisfying assignment.
			Then from each clause $C_i$ for $i\in [m]$, there is a literal $l_{i_j}$ which is set to
			true in $\sigma$.  We take all vertices of $G$ of the form $v^i_{l_j}$ and
			add it to $K$. The subgraph $G[K]$ forms a
			clique of size~$m$. 
			On the other hand suppose we have an $m$-clique
			$K_m$ as a subgraph of $G$. Then, clearly from each partition $p_i$, there exists
			exactly one vertex $v^i_{l_j}$ which is in $K_m$. We set the literal
			$l_j$ to true. This gives a valid satisfying assignment for
			$C$.
		\end{ClaimProof}
	\end{proof}
	
	Now we trim $H$ to avoid the above problem. Again we will use the
	special properties of the touching relation of blocks. We say that some edge
	$e\in E(H)$ is \emph{long}, if one end of $e$ is in $\BTSset{b_i}$, and the
	other in block type $\BTSset{b_j}$ where $j>i+1$. The
	\emph{length} of $e$ is $j-i$. \emph{Delete} all long edges from $H$ to
	obtain the graph $R_H$. In other words we can construct $R_H$ directly,
	similar to $H$, without adding long edges. In the following we first prove that in
	linear time we can construct the graph $R_H$. Second we show that if
	there is an independent set $I$ of size exactly $\sizeof{\blockset}$ in $R_H$
	then $I$ is also an independent set of $H$.
	\begin{lemma}\label[lemma]{lem:computingrh}
		There is an algorithm which computes $R_H$ in time $O((k\cdot k!)^2\sizeof{G})$.
	\end{lemma}
	\begin{proof}
		The algorithm is similar to the construction of $H$. For completeness
		we repeat it here and then we prove it takes time proportional to
		$(k\cdot k!)^2\sizeof{G}$.
		
		\begin{algorithm}\textbf{Construction of $R_H$}\label[algorithm]{alg:rh} 
			\begin{enumerate}[i]
				\item [] \textbf{Input: Update Flow Network $G$}
				\item Set $H\coloneqq \emptyset$, $\blockset':=\blockset$, $\BTSsets:=\emptyset$.
				\item For $i\coloneqq 1,\ldots,\sizeof{\blockset}$ do
				\begin{enumerate}[1]
					\item Let $b\coloneqq b_{\sizeof{\blockset}-i+1}$.
					\item Let $\TBs{b}\coloneqq \{b'_1,\ldots,b'_t\}$ be the set of blocks in $\blockset'$ 
					which touch $b$.
					\item \label{algcomp:permutations}Let $\pi\coloneqq \{\pi_1,\ldots,\pi_{\ell}\}$ be the 
					set of congestion
					free permutations of $\TBs{b}$, compute $\pi$ by the
					algorithm provided in~\ref{lem:computingcongestionfree}.
					\item Set $\BTSset{b}\coloneqq \emptyset$.
					\item For $i\in[\ell]$ create a vertex $v_{\pi_i}$ with
					$\Label(v_{\pi_i})=\pi_i$ and set $\BTSset{b}\coloneqq \BTSset{b}\cup v_{\pi_i}$.
					\item Set $H\coloneqq H\cup \BTSset{b}$.
					\item Add edges between all pairs of vertices in $H[\BTSset{b}]$.
					\item \label{algcomp:edges}Add an edge between every pair of vertices $v\in 
					H[\BTSset{b}]$ and $u\in
					\BTSset{b_{\sizeof{\blockset}-i+2}}$ if the labels of $v$ and $u$ are
					inconsistent and if $b_{\sizeof{\blockset}-i+2}$ exists.
					\item \label{def:removeb} Set $\blockset':=\blockset' - b$.
				\end{enumerate}
			\end{enumerate}
		\end{algorithm}
		The only difference between the above algorithm and the construction of
		$H$ is line~\ref{algcomp:edges}, where we add at most
		$O(k!^2)$ edges to the graph. As there are at most $\sizeof{\blockset}$ steps
		in the algorithm, this shows that the size of $R_H$ is at most
		$O(\sizeof{\blockset}\cdot k!^2)$. Moreover, as there are at most $O(k\sizeof{E(G)})$
		blocks in $G$, the total size of $R_H$ w.r.t.~$G$ is at most
		$O(k\cdot k!^2\cdot \sizeof{G})$. The computations in all other lines except for
		line~\ref{algcomp:permutations} are linear in $k$, hence we only show that the
		total amount of computations in
		line~\ref{algcomp:permutations} is in $O(k!\cdot \sizeof{G})$. We know
		that every edge appears in at most $k$ blocks, hence the
		algorithm provided in~\Cref{lem:computingcongestionfree}, for each edge,
		runs at most $k$ times and as per individual round of that algorithm,
		takes $O(k\cdot \sizeof{G})$. Since there are $k!$ possible permutations
		for each block, this yields a running time of $O(k^2\cdot
		k!\cdot \sizeof{G})$. So all in all, the construction of $R_H$ takes at most
		$O((k\cdot k!)^2\sizeof{G})$ operations.
	\end{proof}
	
	In the above lemma note that we can run the algorithm in parallel. Hence using parallelization,
	the algorithm could be sped up in practice.
	
	\begin{lemma}\label[lemma]{lem:rhandh}
		$H$ has an independent set $I$ of size $\sizeof{\blockset}$ if, and only if, $I$ is also an
		independent set of size $\sizeof{\blockset}$ in $R_H$.
	\end{lemma}
	\begin{proof}
		One direction is clear: if $I$ is an independent set of size $\sizeof{\blockset}$ in
		$H$, then it is an independent set of size $\sizeof{\blockset}$ in $R_H$. On the other hand,
		suppose $I$ is an independent set of size $\sizeof{\blockset}$ in $R_H$. Then for the sake of
		contradiction, suppose there are vertices $u,v\in I$ and an edge
		$e=\{u,v\}\in E(H)$, where $e$ has the shortest length among all
		possible long edges in $H[I]$. 
		Let us assume that $u\in \BTSset{b_i},v\in \BTSset{b_j}$ where
		$j>i+1$. Suppose from each $\BTSset{b_\ell}$ for $i\le \ell \le j$, we
		have $v_{b_\ell}\in I$, where $v_{b_i}=u,v_{b_j}=v$. Clearly as $I$ is
		of size $\sizeof{\blockset}$ there should be exactly one vertex from each
		$\BTSset{b_\ell}$. We know $core(\Label(u),\Label(v))\neq \emptyset$ as otherwise
		the edge $e=\{u,v\}$ was not in $E(H)$. On the other hand, as $e$ is the smallest
		long edge which connects vertices of $I$, then there is no long edge
		between $v_{b_i}$ and $v_{b_{j-1}}$ in $H$. That means $\Label{(v_{b_i})}\consistent 
		\Label{(v_{b_{j-1}})}$
		but then as $\Label{(v_{b_i})}\not\consistent \Label{(v_{b_j})}$
		and by~\nameref{lem:touchesomethinginmiddle} we know that
		$core(\Label(u),\Label(v))\subseteq \Label(v_{b_{j-1}})$, so
		$\Label(v_{b_j})\not\consistent \Label(v_{b_{j-1}})$. Therefore, there is an edge
		between $v_{b_j}$ and $v_{b_{j-1}}$: 
		a contradiction, by our choice of $I$ in $R_h$.
	\end{proof}
	
	$R_H$ is a much simpler graph compared to $H$, which helps us find
	a large independent set of size $\sizeof{\blockset}$ (if exists).
	We have the following lemma.
	
	\begin{lemma}\label[lemma]{lem:computingrh}
		There is an algorithm that finds an
		independent set $I$ of size exactly $\sizeof{\blockset}$ in $R_H$ if such an
		independent set exists; otherwise it outputs that there is no such an
		independent set. The running time of this algorithm is $O(\sizeof{R_H})$.
	\end{lemma}
	\begin{proof}
		We find an independent set of size $\sizeof{\blockset}$ (or we output
		there is no such set) by dynamic programming.
		For this purpose we define a function $f:[\sizeof{\blockset}]\times V(R_H)\rightarrow
		2^{V(R_H)}$ which is presented in detail in the algorithm below. Before
		providing said algorithm we explain it in plain text. It is a straightforward dynamic program:
		start from the left most groups of vertices in $R_H$ (one extreme
		side of $R_H$). Consider
		every vertex as part of the independent set and build the
		independent set bottom up on those groups. We omit the proof of
		correctness and the exact calculation of the running time as
		it is clear from the algorithm.
		
		\begin{algorithm}\textbf{Finding an Independent Set of Size
				$\sizeof{\blockset}$ in $R_H$}\label[algorithm]{alg:findindependentset} 
			\begin{enumerate}
				\item [] \textbf{Input: $R_H$}
				\begin{enumerate}
					\item Set $f(i,v):=\emptyset$ for all $i\in [\sizeof{\blockset}],v\in V(R_H)$.
					\item Set $f(1,v):=v$ for all $v\in \BTSset{b_1}$.
					\item For $2\le i \le [\sizeof{\blockset}]$ do
					\begin{enumerate}
						\item For all $v\in \BTSset{b_i}$ 
						\begin{enumerate}
							\item If there is a vertex $u\in \BTSset{b_{i-1}}$ and $\sizeof{ f(i-1,u)} = i-1$ and 
							$\{u,v\}\not\in
							E(R_H)$ then $f(i,v):=f(i-1,u)\cup \{v\}$,
							\item otherwise set $f(i,v):=\emptyset$
						\end{enumerate}
					\end{enumerate}
					\item If $\exists v\in
					\BTSset{b_{\sizeof{\blockset}}}$ where
					$\sizeof{f(\sizeof{\blockset},v)}=\sizeof{\blockset}$ then output
					$f(\sizeof{\blockset},v)$,
					\item otherwise output there is no such independent set.
				\end{enumerate}
			\end{enumerate}
		\end{algorithm}
	\end{proof}
	Our main theorem is now a corollary of
	the previous lemmas and algorithms.
	\begin{theorem}
		There is a linear time FPT algorithm for the
		network update problem on an acyclic update flow network $G$
		with $k$ flows (the parameter), which finds a
		feasible update sequence, if it exists; otherwise it outputs that there
		is no feasible solution for the given instance. The algorithm
		runs in time $O(2^{O(k\log k)}\sizeof{G})$.
	\end{theorem}
	\begin{proof}
		First construct $R_H$ using \Cref{alg:rh}, then find the
		independent set $I$ of size $\sizeof{\blockset}$ in $R_H$ using
		\Cref{alg:findindependentset}. If there is no such
		independent set $I$, then we output that there is no feasible update
		solution for the given network; this is a consequence
		of~\Cref{lem:rhandh,lem:blocksandsets}. On the other hand, if there is
		such an independent set $I$, then one can construct the corresponding
		dependency graph and update all blocks, using the algorithm provided in
		the proof of~\Cref{lem:blocksandsets}. The dominant runtime term in the
		above algorithms is $O(k^2\cdot k!^2\cdot \sizeof{G})$ (from~\Cref{lem:computingrh}), 
		which proves the
		claim of the theorem.
	\end{proof}

\subsection{Updating $k$-Flows in DAGs is NP-complete}\label{k-complete}
	
	In this section we show that, if the number of flows, $k$, is part of the input,
	the problem remains
	hard even on DAGs. In fact, we prove the following theorem.
	
	\begin{theorem}\label[theorem]{thm:hardondag}
		Finding a feasible update sequence for 
		$k$-flows is NP-complete, even if the update 
		graph $G$ is acyclic.
	\end{theorem}
		
	To prove the theorem, we provide a polynomial time reduction from
	the $3$-SAT problem. Let $C=C_1\wedge\ldots\wedge C_m$ be an instance of
	$3$-SAT with $n$ variables $X_1,\ldots,X_n$, where each variable $X_i$
	appears positive ($x_i$) or negative ($\bar{x}_i$) in some clause $C_j$. 
	We construct an acyclic network update graph $G$ such that there is a
	feasible sequence of updates $\updatesequence$ for
	$G$, if and only if $C$ is satisfiable by some variable assignment
	$\sigma$. By~\Cref{lem:updatewholeblock}, we know 
	that if $G$ has a feasible update sequence, then there is a feasible update
	sequence which updates each block in consecutive rounds.
	
	In the following, we denote the first vertex of a directed path $p$ with
	$head(p)$ and the end vertex with $tail(p)$. Furthermore, we number
	the vertices of a path $p$ with numbers $1,\ldots, \sizeof{ V(p)}$, according to their
	order of appearance in $p$ ($head(p)$ is number $1$). We will write
	$p(i)$ to denote the $i$'th vertex in $p$.
	
	We now describe how to construct the initial update flow network $G$. 
	
	\begin{enumerate}
		\item $G$ has a start vertex $s$ and a terminal vertex $t$.
		
		\item We define $n$ variable selector flow pairs $S_1,\ldots,S_n$, where
		each $S_i=(S^o_i,S^u_i)$ is of demand $1$, as
		follows:
		\begin{enumerate}
			\item \textbf{Variable Selector Old Flows} are $n$  $s,t$-flows $S^o_1,\ldots,S^o_n$
			defined as follows:
			Each one consists of a directed path of length~$3$, where every
			edge in path $S^o_i$ (for $i\in [n]$) has capacity~$1$, except for the edge 
			$(S^o_i(2),S^o_i(3))$,
			which has capacity~$2$. 
			
			\item \textbf{Variable Selector Update Flows } are $n$ $s,t$-flows $S^u_1,\ldots,S^u_n$
			defined as follows:
			Each consists of a directed path of length~$5$, where the
			edge's capacity of path $S^u_i$ is set as follows. $(S^u_i(2),S^u_i(3))$
			has capacity~$2$, $(S^u_i(4),S^u_i(5))$ has capacity~$m$, and the rest
			of its edges has capacity~$1$.
		\end{enumerate}
		
		\item We define $m$ clause flow pairs $C_1,\ldots,C_n$, where
		each $C_i=(C^o_i,C^u_i)$ is of demand $1$, as follows.
		
		\begin{enumerate}
			
			\item\textbf{Clauses Old Flows } are $m$ $s,t$-flows $C^o_1,\ldots,C^o_m$,
			each of length~$5$, where for $i,j\in [m]$, $C^o_i(3)=C^o_j(3)$ and $C^o_i(4)=C^o_j(4)$.  
			Otherwise they are disjoint from the above defined. 
			The edge
			$(C^o_i(3),C^o_i(4))$ (for $i\in [m]$) has capacity~$m$, all other
			edges in $C^o_i$ have capacity~$1$.
			
			\item \textbf{Clauses Update Flows } are $m$ $s,t$-flows 
			$C^u_1,\ldots,C^u_m$, each of length~$3$. Every edge in those paths has
			capacity~$3$.
		\end{enumerate}
		
		\item We define a Clause Validator flow pair $V=(V^o,V^u)$ of demand $m$, as follows.
		\begin{enumerate}
			\item \textbf{Clause Validator Old Flow } is an $s,t$-flow $V^o$ whose
			path consists of edges 
			$(s,S^u_1(4)),S^u_i(4),S^u_i(5)),(S^u_i(5),S^u_{i+1}(4)),(S^u_n(4),S^u_n(5)),(S^u_n(5),t)$
			for $i\in[n-1]$.
			Note that, the edge $(S^u_i(4),S^u_i(5))$ (for $i\in[n]$) also belongs
			to $S^u_i$. All edges of $V$ have capacity~$m$.
			
			\item \textbf{Clause Validator Update Flow } is an $s,t$-flow $V^u$
			whose path has length~$3$, such that
			$V^u(2)=C^o_1(3),V^u(3)=C^o_1(4)$. All new edges of $V^u$ have
			capacity~$m$.
		\end{enumerate}
		
		\item We define $2n$ literal flow pairs $L_1,\ldots,L_{2n}$. 
		Each $L_i=(L^o_i,L^u_i)$ of demand $1$ is defined as follows:
		
		\begin{enumerate}	
			\item\textbf{Literal's Old Flows } are $2n$ $s,t$-flows
			$L^o_1,\ldots,L^o_{n}$ and $\bar{L}^o_1,\ldots,\bar{L}^o_{n}$. Suppose $x_i$ appears in 
			clauses
			$C_{i_1},\ldots,C_{i_\ell}$, then the path $L^o_i$ is a path of length
			$2\ell + 5$, where $L^o_i(2j+1)=C^u_{i_j}(2),L^o_i(2j+2)=C^u_{i_j}(3)$ for $j\in[\ell]$ and
			furthermore $L^o_i(2\ell+3)=S^u_i(2),L^o_i(2\ell+4)=S^u_i(3)$. On
			the other hand, if $\bar{x}_i$ appears in clauses
			$C_{i_1},\ldots,C_{i_{\ell'}}$, then $\bar{L}^o_i$ is a path of
			length $2\ell'+5$ where $\bar{L}^o_i(2j+3)=C^u_{i_j}(,\bar{L}^o_i(2j+4)=C^u_{i_j}(3)$ for 
			$j\in[\ell']$, and
			furthermore
			$\bar{L}^o_i(2\ell'+3)=S^u_i(2),\bar{L}^o_i(2\ell'+4)=S^u_i(3)$. 
			All new edges in $L^o_i$ (resp.~$\bar{L}^o_i$) have capacity~$3$. Note
			that some $L^o_i$s may share common edges.
			
			\item \textbf{Literal's Update Flows } are $2n$ $s,t$-flows
			$L^u_1,\ldots,L^u_{n}$ and $\bar{L}^u_1,\ldots,\bar{L}^u_n$. 
			For $i\in [n]$, $L^u_i$ and $\bar{L}^u_i$ are paths of length~$5$
			such that $L^u_i(2)=\bar{L}^u_i(2)=S^o_i(2)$ and
			$L^u_i(3)=\bar{L}^u_i(3)=S^o_i(3)$. All new edges in those paths
			have capacity~$3$.
		\end{enumerate}
	\end{enumerate}
	
	\begin{figure}[h!]
		\begin{center}
			\begin{tikzpicture}[scale=0.6]
			\tikzset{>=latex} 
			
			\node (s-1) [inner sep=0pt] {};		
			\node (t-1) [inner sep=0pt,position=0:150mm from s-1] {};
			
			\foreach\i in {2,...,13} {
				\pgfmathtruncatemacro{\iPrec}{\i-1};
				\node (s-\i) [inner sep=0pt,position=270:15mm from s-\iPrec] {};
				\node (t-\i) [inner sep=0pt,position=270:15mm from t-\iPrec] {};			
			}
			
			\node (s) [position=260:16mm from s-6,draw,minimum height=120mm,rounded 
			corners=1.5mm,minimum width=3mm,thick] {};
			\node (t) [position=280:16mm from t-6,draw,minimum height=120mm,rounded 
			corners=1.5mm,minimum width=3mm,thick] {};
			
			\node (Liu-4) [draw,circle,thick,position=180:33mm from t-1,scale=0.8] {};	
			\node (Liu-5) [draw,circle,thick,position=180:16.5mm from t-1,scale=0.8] {};

			\node (Sio-2) [draw,circle,thick,position=0:49mm from s-2,scale=0.8] {};
			\node (Sio-3) [draw,circle,thick,position=180:49mm from t-2,scale=0.8] {};	
			\node (Sio-2-o) [position=90:1.3mm from Sio-2,scale=0.8] {};
			\node (Sio-3-o) [position=90:1.3mm from Sio-3,scale=0.8] {};	
			\node (Sio-2-u) [position=270:1.3mm from Sio-2,scale=0.8] {};
			\node (Sio-3-u) [position=270:1.3mm from Sio-3,scale=0.8] {};

			\node (Liub-4) [draw,circle,thick,position=180:33mm from t-3,scale=0.8] {};	
			\node (Liub-5) [draw,circle,thick,position=180:16.5mm from t-3,scale=0.8] {};

			\node (Cko-2) [draw,circle,thick,position=0:26mm from s-5,scale=0.8] {};	
			\node (Cko-3) [draw,circle,thick,position=0:52.5mm from s-5,scale=0.8] {};	
			\node (Cko-5) [draw,circle,thick,position=180:26mm from t-5,scale=0.8] {};	
			\node (Cko-4) [draw,circle,thick,position=180:52mm from t-5,scale=0.8] {};	
			\node (Cko-3-o) [position=90:1.3mm from Cko-3,scale=0.8] {};	
			\node (Cko-4-o) [position=90:1.3mm from Cko-4,scale=0.8] {};
			\node (Cko-3-u) [position=270:1.3mm from Cko-3,scale=0.8] {};	
			\node (Cko-4-u) [position=270:1.3mm from Cko-4,scale=0.8] {};

			\node (Ck'o-2) [draw,circle,thick,position=0:40mm from s-6,scale=0.6] {};	
			\node (Ck'o-5) [draw,circle,thick,position=180:40mm from t-6,scale=0.6] {};
			\node (6-1) [position=205:15mm from Ck'o-2,inner sep=0] {};	
			\node (6-2) [position=335:15mm from Ck'o-5,inner sep=0] {};

			\node (Cku-2) [draw,circle,thick,position=0:49mm from s-7,scale=0.8] {};
			\node (Cku-3) [draw,circle,thick,position=180:49mm from t-7,scale=0.8] {};
			\node (Cku-2-o) [position=90:1.3mm from Cku-2,scale=0.8] {};	
			\node (Cku-3-o) [position=90:1.3mm from Cku-3,scale=0.8] {};
			\node (Cku-2-u) [position=270:1.3mm from Cku-2,scale=0.8] {};	
			\node (Cku-3-u) [position=270:1.3mm from Cku-3,scale=0.8] {};

			\node (Chu-3) [draw,circle,thick,position=0:40mm from s-8,scale=0.8] {};	
			\node (Ch'u-2) [draw,circle,thick,position=180:40mm from t-8,scale=0.8] {};	
			\node (8-1) [position=0:26mm from s-8,inner sep=0] {};	
			\node (8-2) [position=315:15mm from Ch'u-2,inner sep=0] {};

			\node (Cpu-3) [draw,circle,thick,position=0:49mm from s-9,scale=0.6] {};	
			\node (Cp'u-2) [draw,circle,thick,position=180:49mm from t-9,scale=0.6] {};	
			\node (9-1) [position=225:15mm from Cpu-3,inner sep=0] {};	
			\node (9-2) [position=315:15mm from Cp'u-2,inner sep=0] {};

			\node (Lio-2l+2) [draw,circle,thick,position=0:20mm from s-10,scale=0.8] {};	
			\node (Lio-2l+5) [draw,circle,thick,position=0:55mm from s-10,scale=0.8] {};	
			\node (10-1) [position=90:15mm from Lio-2l+2,inner sep=0] {};

			\node (Liob-2l+2) [draw,circle,thick,position=0:10mm from s-11,scale=0.6] {};	
			\node (Liob-2l+5) [draw,circle,thick,position=0:65mm from s-11,scale=0.6] {};
			\node (11-1) [position=90:15mm from Liob-2l+2,inner sep=0] {};

			\node (Siu-2) [draw,circle,thick,position=0:26mm from s-12,scale=0.8] {};	
			\node (Siu-3) [draw,circle,thick,position=0:52.5mm from s-12,scale=0.8] {};	
			\node (Siu-5) [draw,circle,thick,position=180:26mm from t-12,scale=0.8] {};	
			\node (Siu-4) [draw,circle,thick,position=180:52mm from t-12,scale=0.8] {};
			\node (Siu-2-o) [position=90:1.3mm from Siu-2,scale=0.8] {};	
			\node (Siu-3-o) [position=90:1.3mm from Siu-3,scale=0.8] {};
			\node (Siu-2-u) [position=270:1.3mm from Siu-2,scale=0.8] {};	
			\node (Siu-3-u) [position=270:1.3mm from Siu-3,scale=0.8] {};	
			\node (Siu-4-u) [position=270:1.3mm from Siu-4,scale=0.8] {};	
			\node (Siu-5-u) [position=270:1.3mm from Siu-5,scale=0.8] {};

			\node (S1u-4) [draw,circle,thick,position=0:15.5mm from s-13,scale=0.8] {};	
			\node (Si-1u-5) [draw,circle,thick,position=180:60mm from t-13,scale=0.8] {};	
			\node (Si+1u-4) [draw,circle,thick,position=180:15mm from t-13,scale=0.8] {};
			\node (13-1) [position=0:25.5mm from s-13,inner sep=0] {};	
			\node (13-2) [position=0:50mm from s-13,inner sep=0] {};	
			\node (13-3) [position=180:8mm from t-13,inner sep=0] {};

			\draw [color=Dandelion,line width=1.3pt,->,dashed] (s-1) to (Sio-2);
			\draw [color=Dandelion,line width=1.3pt,->,dashed] (Sio-2-o) to (Sio-3-o); 
			\draw [color=Dandelion,line width=1.3pt,->,dashed] (Sio-3) to (Liu-4); 
			\draw [color=Dandelion,line width=1.3pt,->,dashed] (Liu-4) to (Liu-5); 
			\draw [color=Dandelion,line width=1.3pt,->,dashed] (Liu-5) to (t-1); 
			
			\draw [color=blue,line width=1.3pt,->] (s-2) to (Sio-2);
			\draw [color=blue,line width=1.3pt,->] (Sio-2) to (Sio-3); 
			\draw [color=blue,line width=1.3pt,->] (Sio-3) to (t-2); 		
			
			\draw [color=Dandelion,line width=1.3pt,->,dash pattern=on 2pt off 5.5pt, dash phase=2pt] 
			(s-3) to (Sio-2);
			\draw [color=Dandelion,line width=1.3pt,->,dash pattern=on 2pt off 5.5pt, dash phase=2pt] 
			(Sio-2-u) to (Sio-3-u); 
			\draw [color=Dandelion,line width=1.3pt,->,dash pattern=on 2pt off 5.5pt, dash phase=2pt] 
			(Sio-3) to (Liub-4); 
			\draw [color=Dandelion,line width=1.3pt,->,dash pattern=on 2pt off 5.5pt, dash phase=2pt] 
			(Liub-4) to (Liub-5); 
			\draw [color=Dandelion,line width=1.3pt,->,dash pattern=on 2pt off 5.5pt, dash phase=2pt] 
			(Liub-5) to (t-3); 
			
			\draw [color=Red,line width=2.3pt,->,dashed] (s-4) to (Cko-3);
			\draw [color=Red,line width=2.3pt,->,dashed] (Cko-3-o) to (Cko-4-o); 
			\draw [color=Red,line width=2.3pt,->,dashed] (Cko-4) to (t-4); 
			
			\draw [color=JungleGreen,line width=1.3pt,->] (s-5) to (Cko-2);	
			\draw [color=JungleGreen,line width=1.3pt,->] (Cko-2) to (Cko-3);	
			\draw [color=JungleGreen,line width=1.3pt,->] (Cko-3) to (Cko-4);	
			\draw [color=JungleGreen,line width=1.3pt,->] (Cko-4) to (Cko-5);	
			\draw [color=JungleGreen,line width=1.3pt,->] (Cko-5) to (t-5);	
			
			\draw [color=JungleGreen,line width=0.9pt,->,decoration = {zigzag,segment length = 2mm, 
			amplitude = 0.3mm}, decorate] (6-1) to (Ck'o-2);	
			\draw [color=JungleGreen,line width=0.9pt,->,decoration = {zigzag,segment length = 2mm, 
			amplitude = 0.3mm}, decorate] (Ck'o-2) to (Cko-3);	
			\draw [color=JungleGreen,line width=0.9pt,->,decoration = {zigzag,segment length = 2mm, 
			amplitude = 0.3mm}, decorate] (Cko-3-u) to (Cko-4-u);	
			\draw [color=JungleGreen,line width=0.9pt,->,decoration = {zigzag,segment length = 2mm, 
			amplitude = 0.3mm}, decorate] (Cko-4) to (Ck'o-5);	
			\draw [color=JungleGreen,line width=0.9pt,->,decoration = {zigzag,segment length = 2mm, 
			amplitude = 0.3mm}, decorate] (Ck'o-5) to (6-2);		
			
			\draw [color=JungleGreen,line width=1.3pt,->,dashed] (s-7) to (Cku-2);
			\draw [color=JungleGreen,line width=1.3pt,->,dashed] (Cku-2-o) to (Cku-3-o); 
			\draw [color=JungleGreen,line width=1.3pt,->,dashed] (Cku-3) to (t-7); 
			
			\draw [color=Dandelion,line width=1.3pt, dash pattern=on .13pt off 5pt, dash phase=2pt, 
			line cap=round] (s-8) to (8-1);
			\draw [color=Dandelion,line width=1.3pt,->] (8-1) to (Chu-3);	
			\draw [color=Dandelion,line width=1.3pt,->] (Chu-3) to (Cku-2);	
			\draw [color=Dandelion,line width=1.3pt,->] (Cku-2) to (Cku-3);	
			\draw [color=Dandelion,line width=1.3pt,->] (Cku-3) to (Ch'u-2);	
			\draw [color=Dandelion,line width=1.3pt,->] (Ch'u-2) to (8-2);			 
			
			\draw [color=Dandelion,line width=0.9pt,->,decoration = {zigzag,segment length = 2mm, 
			amplitude = 0.3mm}, decorate] (9-1) to (Cpu-3);	
			\draw [color=Dandelion,line width=0.9pt,->,decoration = {zigzag,segment length = 2mm, 
			amplitude = 0.3mm}, decorate] (Cpu-3) to (Cku-2);	
			\draw [color=Dandelion,line width=0.9pt,->,decoration = {zigzag,segment length = 2mm, 
			amplitude = 0.3mm}, decorate] (Cku-2-u) to (Cku-3-u);	
			\draw [color=Dandelion,line width=0.9pt,->,decoration = {zigzag,segment length = 2mm, 
			amplitude = 0.3mm}, decorate] (Cku-3) to (Cp'u-2);	
			\draw [color=Dandelion,line width=0.9pt,->,decoration = {zigzag,segment length = 2mm, 
			amplitude = 0.3mm}, decorate] (Cp'u-2) to (9-2);	
			
			\draw [color=Dandelion,line width=1.3pt,->] (10-1) to (Lio-2l+2);	
			\draw [color=Dandelion,line width=1.3pt,->] (Lio-2l+2) to (Siu-2);	
			\draw [color=Dandelion,line width=1.3pt,->] (Siu-2-o) to (Siu-3-o);	
			\draw [color=Dandelion,line width=1.3pt,->] (Siu-3) to (Lio-2l+5);	
			\draw [color=Dandelion,line width=1.3pt,->] (Lio-2l+5) to (t-10);
			
			\draw [color=Dandelion,line width=0.9pt,->] (11-1) to (Liob-2l+2);	
			\draw [color=Dandelion,line width=0.9pt,->] (Liob-2l+2) to (Siu-2);	
			\draw [color=Dandelion,line width=0.9pt,->] (Siu-2) to (Siu-3);	
			\draw [color=Dandelion,line width=0.9pt,->] (Siu-3) to (Liob-2l+5);	
			\draw [color=Dandelion,line width=0.9pt,->] (Liob-2l+5) to (t-11);
			
			\draw [color=blue,line width=1.3pt,->,dashed] (s-12) to (Siu-2);	
			\draw [color=blue,line width=1.3pt,->,dashed] (Siu-2-u) to (Siu-3-u);	
			\draw [color=blue,line width=1.3pt,->,dashed] (Siu-3) to (Siu-4);	
			\draw [color=blue,line width=1.3pt,->,dashed] (Siu-4) to (Siu-5);	
			\draw [color=blue,line width=1.3pt,->,dashed] (Siu-5) to (t-12);
			
			\draw [color=Red,line width=2.3pt,->] (s-13) to (S1u-4);	
			\draw [color=Red,line width=2.3pt,->] (S1u-4) to (13-1);	
			\draw [color=Red,line width=2.3pt,dash pattern=on .13pt off 5pt, dash phase=2pt, line 
			cap=round] (13-1) to (13-2);	
			\draw [color=Red,line width=2.3pt,->] (13-2) to (Si-1u-5);	
			\draw [color=Red,line width=2.3pt,->] (Si-1u-5) to (Siu-4);
			\draw [color=Red,line width=2.3pt,->] (Siu-4-u) to (Siu-5-u);	
			\draw [color=Red,line width=2.3pt,->] (Siu-5) to (Si+1u-4);	
			\draw [color=Red,line width=1.3pt,->] (Si+1u-4) to (13-3);	
			\draw [color=Red,line width=2.3pt, dash pattern=on .13pt off 5pt, dash phase=2pt, line 
			cap=round] (13-3) to (t-13);

			\node (Ls) [position=180:7mm from s] {\Large $s$};
			\node (Lt) [position=0:7mm from t] {\Large $t$};

			\node (LSio-2) [position=90:7mm from Sio-2] {$S_i^o(2)$};
			\node (LSio-3) [position=102:7mm from Sio-3] {$S_i^o(3)$};
			\node (LLiu-4) [position=90:7mm from Liu-4] {$L_i^u(4)$};
			\node (LLiu-5) [position=90:7mm from Liu-5] {$L_i^u(5)$};
			\node (LLiub-4) [position=270:7mm from Liub-4] {$\overline{L}_i^u(4)$};
			\node (LLiub-5) [position=270:7mm from Liub-5] {$\overline{L}_i^u(5)$};

			\node (LCko-2) [position=270:7mm from Cko-2] {$C_k^o(2)$};
			\node (LCk'o-2) [position=300:5.5mm from Ck'o-2] {\tiny $C_{k'}^o(2)$};
			\node (LCko-3) [position=90:7mm from Cko-3] {$C_k^o(3)$};
			\node (LCko-4) [position=90:7mm from Cko-4] {$C_k^o(4)$};
			\node (LCk'o-5) [position=270:5mm from Ck'o-5] {\tiny $C_{k'}^o(5)$};
			\node (LCko-5) [position=270:7mm from Cko-5] {$C_k^o(5)$};

			\node (LChu-3) [position=270:7mm from Chu-3] {$C_h^u(3)$};
			\node (LCpu-3) [position=0:8mm from Cpu-3] {\tiny $C_p^u(3)$};
			\node (LCku-2) [position=200:13mm from Cku-2] {$C_k^u(2)$};
			\node (LCku-3) [position=340:12mm from Cku-3] {$C_k^u(3)$};
			\node (LCh'u-2) [position=0:11mm from Ch'u-2] {$C_{h'}^u(2)$};
			\node (LCp'u-2) [position=180:8mm from Cp'u-2] {\tiny $C_{p'}^u(2)$};

			\node (LLio-2l+2) [position=0:15mm from Lio-2l+2] {$L_i^o(2\ell+2)$};
			\node (LLiob-2l+2) [position=270:11mm from Liob-2l+2] {\tiny $\overline{L}_i^o(2\ell+2)$};
			\node (LSiu-2) [position=270:7mm from Siu-2] {$S_i^u(2)$};
			\node (LSiu-3) [position=270:7mm from Siu-3] {$S_i^u(3)$};
			\node (LLio-2l+5) [position=15:15mm from Lio-2l+5] {$L_i^o(2\ell+5)$};
			\node (LLiob-2l+5) [position=15:12mm from Liob-2l+5] {\tiny $\overline{L}_i^o(2\ell+5)$};
			
			\node (LS1u-4) [position=270:7mm from S1u-4] {$S_1^u(4)$};
			\node (LSi-1u-5) [position=270:7mm from Si-1u-5] {$S_{i-1}^u(5)$};
			\node (LSiu-4) [position=90:7mm from Siu-4] {$S_i^u(4)$};
			\node (LSiu-5) [position=90:7mm from Siu-5] {$S_i^u(5)$};
			\node (LSi+1u-4) [position=270:7mm from Si+1u-4] {$S_{i+1}^u(4)$};
			
			\node (gL1) [position=0:74mm from s-2] {};
			\node (gL2) [position=0:39mm from s-12] {};
			\node (gL3) [position=0:110.5mm from s-12] {};
			\node (gL4) [position=0:74mm from s-5] {};
			\node (gL5) [position=0:74mm from s-7] {};
			
			\node (L1) [position=90:5mm from gL1] {$2$};
			\node (L2) [position=90:5mm from gL2] {$2$};
			\node (L3) [position=90:4mm from gL3] {$m$};
			\node (L4) [position=90:5mm from gL4] {$m$};
			\node (L5) [position=90:5mm from gL5] {$3$};
			
			\begin{pgfonlayer}{bg}

			\end{pgfonlayer}

			\end{tikzpicture}	
		\end{center}
		\vspace{-1em}
		\caption{\emph{Gadget Construction for Hardness in DAGs:}
			There are~$4$ types of flows: Clause flows, Literal flows,
			Clause Validator flow and Literal Selector flows. The edge
			$(S_i^o(2),S_i^o(3))$ cannot route~$3$
			different flows~$S_i^o$, $L_i^u$,$\bar{L}_i^u$ at the same
			time. On the
			other hand the edge~$(S_i^u(2),S_i^u(3))$ cannot route the
			flow~$S_i^u$ before updating
			either $L^o_i$ or $\bar{L}^o_i$, hence by the above
			observation, exactly one of the $L_i$ or $\bar{L}_i$'s will be
			updated strictly before $S_i$ and the other will be
			updated strictly after $S_i$ was updated.
			Only after all Clause flows are updated, the
			edge~$(C^o_k(3),C^o_k(4))$ can route the flow~$V$
			(Clause Validator flow). A Clause flow $C_k$ can be updated only if at
			least one of the Literal flows which goes along
			$(C_k^u(2),C_k^u(3))$ is updated. So in each clause, there
			should be a valid literal. On the other hand the Clause
			validator flow can be updated only if all Clause
			Selector flows are updated, this is guaranteed by the edge
			$(S_i^u(4),S_i^u(5))$. Hence, before updating all clauses, we
			are allowed to update at most one of the $L_i$ or
			$\bar{L}_i$'s, and this corresponds to a valid satisfying assignment.
		}
		\label{fig:kflowsarehard}
	\end{figure}
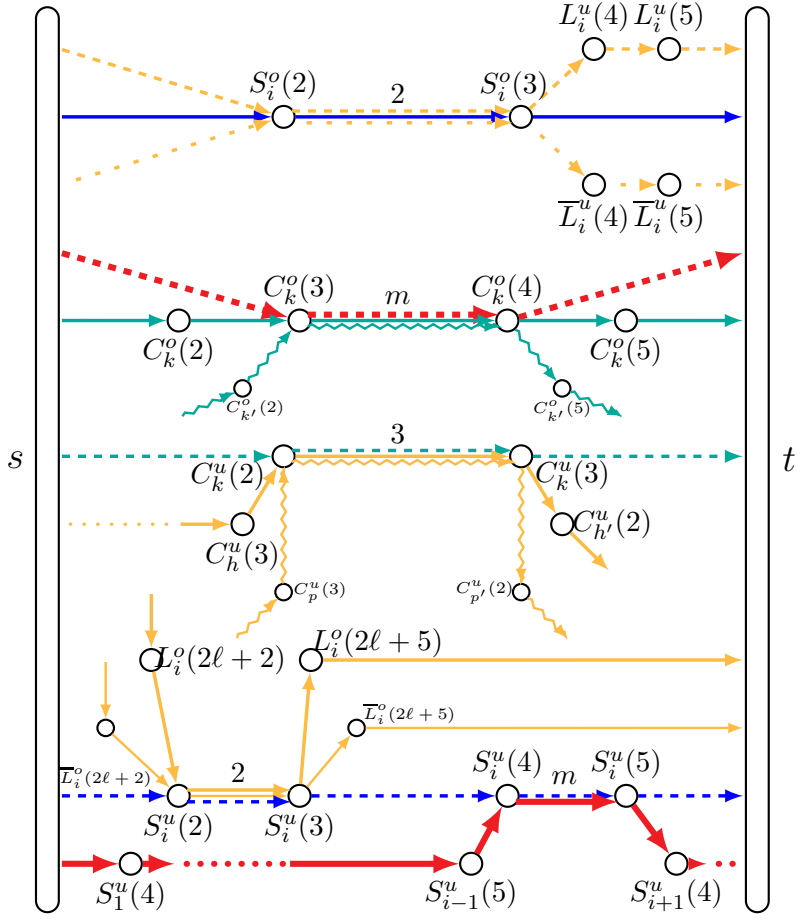

	\begin{lemma}\label[lemma]{lem:allobservations}
		For $\updatesequence$ and $G$, we have the following observations.
		\begin{enumerate}[i)]
			\item \label{obs:switchselector} We either have 
			$\updatesequence(L_i^o)<\updatesequence(S_i^o)<\updatesequence(\bar{L}_i^o)$, or  
			$\updatesequence(\bar{L}_i^o)<\updatesequence(S_i^o)<\updatesequence(L_i^o)$, for all 
			$i\in[n]$ .
			
			\item \label{obs:clausevalidator} $\updatesequence(C_i^o)<\updatesequence(V^o)$ for all 
			$i\in[m]$.
			
			\item \label{obs:clausevalidatorandswitchs} 
			$\updatesequence(S_i^o)<\updatesequence(V^o)$ for all $i\in[n]$.
			
			\item \label{obs:clausevarselector} For every $i\in[m]$ there is some $j\in[n]$ such that 
			$\updatesequence(C_i^o)<\updatesequence(L_j^o)$ or 
			$\updatesequence(C_i^o)<\updatesequence(\bar{L}_j^o)$.
			
			\item \label{obs:clausevar} We either have 
			$\updatesequence(L_j^o)<\updatesequence(C_i^o)<\updatesequence(\bar{L}_j^o)$, or  
			$\updatesequence(\bar{L}_j^o)<\updatesequence(C_i^o)<\updatesequence(L_j^o)$, for all 
			$i\in[m]$ and all $j\in[n]$.
			
		\end{enumerate}
	\end{lemma}
	\begin{proof}~
		\begin{enumerate} [i)]
			\item As the capacity of the edge $e=(S^o_{i}(2),S^o_{i}(3))$ is~$2$, and both
			$L^u_i,\bar{L}^u_i$ use that edge, before updating both of
			them, $S^o_i$ (resp. $S^u_i$) should be updated. On the other hand, the edge
			$e'=(S^u_{i}(2),S^u_{i}(3))$ has capacity~$2$ and it is in both
			$L^o_i$ and $\bar{L}^o_i$. So to update $S^o_{i}$, $e'$ for one of the
			$L^o_i,\bar{L}^o_i$ should be updated.
			
			\item The edge $(V^u(2),V^u(3))$ of $V^u$ also belongs to all $C^o_i$
			(for $i\in[m]$) and its capacity is $m$. Moreover, 
			the demand of $(V^o,V^u)$ is $m$, so $V^o$ cannot be updated 
			unless $C^o_i$ has been updated for all $i\in[m]$.
			
			\item Every $S^u_i$ ($i\in[n]$) requires the edge $(S^u_i(4),S^u_i(5))$, 
			which is also used by $V^o$, until after step $\updatesequence(V^o)$.
			
			\item This is a consequence
			of Observation~\ref{obs:clausevalidatorandswitchs} and
			Observation~\ref{obs:clausevalidator}.
			
			\item This is a consequence of Observation~\ref{obs:clausevarselector} and 
			Observation~\ref{obs:switchselector}.
		\end{enumerate}
	\end{proof}

	\begin{proof}[Proof of~\Cref{thm:hardondag}]
		Given a sequence of updates, we can check if 
		it is feasible or not. The length of the update sequence 
		is at most $k$ times the size of the graph,
		hence, the problem clearly is in NP. 
		
		To show that the problem is complete for NP, 
		we use a reduction from
		$3$-SAT. Let $C$ be as defined earlier in this section, 
		and in polynomial
		time we can construct $G$. 
		
		By the construction of $G$, if there is a satisfying assignment
		$\sigma$ for $C$, we obtain a sequence $\updatesequence$ to update the flows in
		$G$ as follows. First, if in
		$\sigma$ we have $X_i=1$ for some $i\in[n]$, update the
		literal flow $L^o_i$; otherwise update the literal flow
		$\bar{L}^o_i$. Afterwards, since $\sigma$ satisfies $C$, for every clause
		$C_i$ there is some literal flow $L_j$ or $\bar{L}_j$, which is
		already updated. Hence, for all $i\in [m]$ the edge
		$(C^u_i(3),C^u_i(4))$ incurs a load of~$2$ while its capacity is~$3$.
		Therefore, we can update all of the
		clause flows and afterwards the clause validator flow $V^o$. Next, we
		can update the clause selector flows and at the end, we update the
		remaining half of the literal flows.
		These groups of updates can all be done consecutively.
		
		On the other hand, if there is a valid update sequence $\updatesequence$ for
		flows in $G$, by~\Cref{lem:allobservations}
		observation~\ref{obs:clausevar}, there are exactly~$n$ literal flows that
		have to be updated, before we can update $C_i^o$. To be more
		precise, for every $j\in[n]$, either $L_j^o$, or $\bar{L}_j^o$ has to
		be updated, but never both. If $L_j^o$ is one of those first~$n$
		literal flows to be updated for some $j\in[n]$, we set
		$X_j:=1$; otherwise $\bar{L}_j^o$ is to be updated and we
		set $X_j:=0$. Since these choices are guaranteed to be
		unique for every $j\in[n]$, this gives us an assignment
		$\sigma$. After these~$n$ literal flows are updated, we are able to
		update the clause flows, since $\updatesequence$ is a valid update sequence. This
		means in particular, that for every clause $C_i$, $i\in[m]$, there is
		at least one literal which is set to true. Hence $\sigma$ satisfies $C$
		and therefore solving the network update problem on DAGs, is as hard as
		solving the $3$-SAT problem.
	\end{proof}

	\section{Related Work}\label{sec:relwork}
	
	To the best of our knowledge, our model is novel in the context of 
	reconfiguration theory~\cite{van2013complexity}. 
	The reconfiguration model closest to ours is by Bonsma~\cite{bonsma2013complexity} 
	who studied
	how to perform rerouting such that transient paths are always \emph{shortest}.
	However, the corresponding techniques and results are not applicable
	in our model where we consider flows of certain \emph{demands},
	and where different flows may \emph{interfere} due to capacity constraints
	in the underlying network.
	
	The problem of how to update routes of flows has been studied
	intensively by the networking community 
	recently~\cite{infocom15,dionysus,zupdate,roger,abstractions}, 
	in particular in the context of
	software-defined networks and motivated by the unpredictable
	router update times~\cite{dionysus,kuzniar2015you}.
	For an overview,
	we refer the reader to a recent survey by Foerster et al.~\cite{update-survey}.
	In a seminal work by Reitblatt et al.~\cite{abstractions},
	a strong \emph{per-packet consistency} notion has been studied, 
	which is well-aligned with the strong consistency properties
	usually provided in traditional networks~\cite{DBLP:conf/wdag/CernyFJM16}. 
	Mahajan and Wattenhofer~\cite{roger} 
	started exploring the benefits of relaxing the per-packet consistency property, while 
	\emph{transiently} providing only essential properties 
	like loop-freedom.
	The authors also present a first algorithm that
	quickly updates routes in a transiently loop-free manner,
	and their study was recently refined 
	in~\cite{sirocco16update,Forster2016Consistent,Forster2016Power},
	where the authors also establish hardness results,
	as well as in~\cite{dsn16,sigmetrics16,ludwig2015scheduling,hotnets14update},
	which respectively, focus on the problem of minimizing the number
	of scheduling rounds~\cite{ludwig2015scheduling}, 
	initiate the study of multiple policies~\cite{dsn16},
	and introduce additional transient routing constraints 
	related to waypointing~\cite{sigmetrics16,hotnets14update}.
	However, none of these papers considers
	bandwidth capacity constraints.
	
	Congestion is known to 
	negatively affect application performance and
	user experience. 
	The seminal work by Hongqiang et al.~\cite{zupdate} 
	on congestion-free rerouting
	has already been extended in several papers, 
	using static~\cite{roger-infocom,swan,jaq4,icnp-jiaqi}, 
	dynamic~\cite{jaq1}, or time-based~\cite{jaq8,jaq10} 
	approaches. 
	Vissicchio et al.~presented FLIP~\cite{vissicchio2016flip},
	which combines
	per-packet consistent updates with order-based rule replacements, 
	in order to reduce memory overhead:
	additional rules are used only when necessary.
	Moreover, Hua et al.~\cite{huafoum}
	recently initiated the study of 
	adversarial settings,
	and presented FOUM, 
	a flow-ordered
	update mechanism that is robust to packet-tampering and packet dropping
	attacks. 
	
	However, to the best of our knowledge, bandwidth capacity
	constraints have so far only been considered
	in strong, per-packet consistent settings, 
	and for splittable flows. We in this paper argue
	that this is both impractical (splittable flows
	introduce a wide range of problems and overheads)
	as well as too restrictive (per-packet consistent updates require traffic marking and
	render many problem instances infeasible).

	\section{Conclusion}\label{sec:conclusion}
	
	This paper initiated the study of a natural and fundamental
	reconfiguration
	problem: the congestion-free rerouting of unsplittable flows.
	Interestingly, we find that while \emph{computing} disjoint paths on 
	DAGs is $W[1]$-hard~\cite{slivkins2010parameterized} and 
	finding routes under congestion even harder~\cite{AmiriKMR16}, 
	\emph{reconfiguring} multicommodity flows is fixed parameter tractable on DAGs.
	However, we also show that the problem is NP-hard for an arbitrary number of flows.
	
	In future work, it will be interesting
	to chart a more comprehensive landscape of the computational
	complexity for the network update problem.
	In particular, it would be interesting to know whether
	the complexity can be reduced further, e.g., to
	$2^{O(k)}O(\sizeof{G})$. More generally,
	it will be interesting to study other flow graph
	families, especially more sparse graphs or graphs of bounded DAG
	width~\cite{DBLP:journals/tcs/AmiriKR16,DBLP:journals/jct/BerwangerDHKO12}. 
	\bigskip
	
	\noindent \textbf{Acknowledgements.} We would like
	to thank Stephan Kreutzer, Arne Ludwig and Roman Rabinovich for discussions on this problem.
	
	\bibliographystyle{plain}
	\vspace{-5mm}
	{
		\bibliography{literature}
	}

\end{document}